\documentclass[useAMS,usenatbib]{mn2e}
\usepackage{amsfonts,amsmath,amssymb}
\usepackage{graphicx,epsfig}
\usepackage[colour]{optional}

\newcommand{\Msun}{\hbox{$\mathrm{M}_{\odot}$}}

\newcommand{\Feedback}{FO}
\newcommand{\Nocool}{GO}
\newcommand{\Precool}{PC}

\newcommand{\RE}{\mbox{$r_{2500}$}}

\journal{Preprint astro-ph/yymmnnn}

\title[Baryon fractions in clusters]{Baryon fractions in clusters of
  galaxies: evidence against a preheating model for entropy generation}

\author[Young~et~al.]
{Owain E. Young$^1$\thanks{E-mail: o.e.young@sussex.ac.uk},
  Peter A. Thomas$^1$, C. J. Short$^1$ and Frazer Pearce$^2$\\
  {}$^{1}$Astronomy Centre, University of Sussex, Falmer, Brighton BN1 9QH,
  United Kingdom\\
  {}$^{2}$Department of Physics and Astronomy, University of
  Nottingham, Nottingham, NG7 2RD, United Kingdom
}

\begin{document}


\pagerange{\pageref{firstpage}--\pageref{lastpage}} \pubyear{2008}

\maketitle

\label{firstpage}

\begin{abstract}

The Millennium Gas project aims to undertake smoothed-particle
hydrodynamic resimulations of the Millennium Simulation, providing
many hundred massive galaxy clusters for comparison with X-ray surveys
(170 clusters with $kT_\mathrm{sl}>3\,$keV).  This paper looks at the
hot gas and stellar fractions of clusters in simulations with
different physical heating mechanisms.  These fail to reproduce
cool-core systems but are successful in matching the hot gas
profiles of non-cool-core clusters.  Although there is immense
scatter in the observational data, the simulated clusters broadly
match the integrated gas fractions within $r_{500}$.  In line with
previous work, however, they fare much less well when compared to the
stellar fractions, having a dependence on cluster mass that is much
weaker than is observed.  The evolution with redshift of the hot gas
fraction is much larger in the simulation with early preheating than
in one with continual feedback; observations favour the latter model.
The strong dependence of hot gas fraction on cluster physics limits
its use as a probe of cosmological parameters.

\end{abstract}

\begin{keywords}
methods: N-body simulations -- galaxies: clusters: general --
galaxies: evolution
\end{keywords}

\section{Introduction}
\label{sec:intro}

The hot gas fraction of clusters of galaxies was first used as a
cosmological probe by \citet{ASF02}, and later refined in
\citet{ASE04} and \citet{ARS08}.  These papers show quite conclusively
that the gas fraction can be used to derive cosmological parameters
that are in agreement with the concordance $\Lambda$CDM cosmology.  
A similar result was obtained by \citet{LBC06} in a joint X-ray,
Sunyaev-Zel'dovich analysis.  Subsequently, a more sophisticated
analysis that properly takes into account selection effects, and
combines mass, X-ray luminosity and temperature observations of 238
clusters at $z\le0.5$, reached similar conclusions
\citep{MAE09,MAR09}, as did a study by \citet{EMT09} of 60
  clusters extending to $z\sim1.3$.

However, as has been pointed out by \citet{SBV05}, the above
conclusion relies very heavily upon the assumption that the gas
fraction in the clusters used in the study is independent of both mass
and redshift.  We show in this paper that using different models for
entropy generation in the intracluster medium (ICM) can lead to
variations in hot gas fractions in simulated clusters that are at
least as great as those that one obtains by using an incorrect
cosmology in the observational data analysis.  We argue, therefore,
that, at present, it is more useful to fix the cosmology to
the concordance value and to use the data to constrain cluster
physics.  By doing so we conclude that the data favour a model 
of continual energy injection into the ICM from galaxies rather than 
a widespread preheating episode at high redshift.

In the future, once the physical models of the ICM become
  more refined, and with the large statistical samples of clusters
  generated by {\it eROSITA}, it should be possible to do a combined
  analysis that constrains both the cluster physics and the cosmology
  simultaneously.

In Section~\ref{sec:method} we describe the numerical method that we
use, the different feedback schemes, and our method of cluster
identification.  Section~\ref{sec:results} describes our results first
on gas fraction profiles, then scaling relations, and finally the
evolution of each of these.  The conclusions of the paper are
summarised in Section~\ref{sec:conc}.

\section{Method}
\label{sec:method}

Here we present an overview of our numerical scheme.  The method is
described at length in \citet{SHT09} and its particular application to
clusters of galaxies in \citet{STY10}.

\subsection{Simulations}

We present results from three simulations taken from the Millennium
Gas Project, the basic objective of which is to add gas to the dark
matter-only Millennium Simulation \citep{SWJ05}.  Each simulation
incorporates a different model of the baryonic physics, so that we can
assess the impact of varying physical assumptions on the thermal
history of the ICM.

In the first Millennium Gas run, the intracluster gas is heated solely
by gravitational processes. We refer to this simulation as the GO
(Gravitation Only) run. Although this run does not include gas cooling
or heating from astrophysical sources such as supernovae (SNe) and
active galactic nuclei (AGN), it is useful as a base model, enabling
us to determine exactly which cluster properties are affected by
astrophysical processes beyond gravitational heating. Given that the
only source of gas entropy changes in the GO run is gravity, then we
would expect a self-similar cluster population to be formed. This is
generally found to be the case in non-radiative simulations
(e.g. \citealt{NFW95,ENF98,VKB05,ASY06,MKT06,SRE09}).

The second Millennium Gas simulation also includes high-redshift
preheating \citep[first posited by][]{KAI91,EVH91} and
radiative cooling. We name this simulation the PC (Preheating plus
Cooling) run. Preheating raises the entropy of the ICM before
gravitational collapse, preventing gas from reaching high densities in
central cluster regions and thus reducing its X-ray emissivity. This
effect is greater in lower-mass systems, breaking the self-similarity
of the cluster scaling relations in a way that resembles observations
\citep{BEM01,BRM01,MTK02,BGW02,TBS03,BFK05}.

The simple model of preheating employed in the PC simulation is
similar to that of \citet{BGW02}. Briefly, the entropy of every
particle is raised to $200$ keV cm$^2$ at $z=4$, thus creating an
entropy `floor' (note that a particle's entropy is not changed if it
already has a value in excess of this at $z=4$). In
addition to preheating, there is also radiative cooling based on the
cooling function of \citet{SUD93}, assuming a fixed metallicity of
$0.3\,Z_\odot$ (a good approximation to the mean metallicity of the ICM
out to at least $z=1$; \citealt{TRE03}). Once the temperature of a gas
particle drops below $2\times 10^4$ K, the hydrogen density exceeds
$\rho_{\rm H}=4.2\times 10^{-27}$\,g\,cm$^{-3}$ and the density
contrast is greater than $100$, then it is converted to a
collisionless star particle. However, the preheating is so extreme
that star formation is effectively terminated at $z=4$, so that less
than $2\%$ of the baryonic matter is locked-up in stars at $z=0$.
 
Although there is considerable evidence that non-gravitational heating
of the ICM indeed occurs mainly at high redshift
(e.g.~\citealt{EBG08,WCP09}), the preheating scenario is clearly a
gross simplification of the complex interplay between star formation,
black hole growth and associated feedback. Despite this, preheating
does provide a useful effective model for the effects of
non-gravitational heating. In particular, the Millennium Gas PC run
can reproduce several key observational properties of the low-redshift
cluster population, including halo gas fractions \citep{SRE09}.
However, the model fails to account for the observed scatter about the
mean relations, particularly on group scales, and generates over-large
isentropic cores in low-mass systems as compared to observational data
(e.g.~\citealt{PSF03,PAP06}).

Finally, we use a recent addition to the Millennium Gas suite in which
feedback is directly tied to galaxy formation, rather than assuming
some \emph{ad hoc} injection of energy at high-redshift. We term this
the FO (Feedback Only) run to emphasise the fact that the model
currently does not include radiative cooling.  The model we use is the
hybrid scheme of \citet{SHT09}, where a semi-analytic model is used to
calculate the energy transferred to the intracluster gas by SNe and
AGN. An immediate benefit of this approach is that feedback is
guaranteed to originate from a galaxy population whose observational
properties agree well with those of real galaxies. This is generally
not the case in fully self-consistent hydrodynamical simulations that
include radiative cooling and stellar feedback because too much gas
cools out of the hot phase, leading to excessive star formation
(e.g. \citealt{BMS04,KSA07}). It is widely thought that additional
heating from AGN is the natural solution to this overcooling
problem. Indeed, \citet{CSP09} and \citet{FBT10} have demonstrated
that including AGN feedback in hydrodynamical simulations can
successfully balance radiative cooling in galaxy groups. However, the
stellar fraction is still found to be 2 to 3 times larger than
observed in massive clusters.

For reasons of computational efficiency, the FO run was not undertaken
over the entire Millennium Simulation volume.  Instead, we resimulated
a sample of several hundred galaxy groups and clusters.  Each run came
in three distinct stages: a dark matter-only resimulation of each
region containing a cluster from our sample; semi-analytic galaxy
catalogues built on the halo merger trees of these resimulations; and
hydrodynamical resimulations of the same regions to track the energy
injection from model galaxies.  The merger trees were built using the
procedure of \citet{SWJ05} and we used the Munich L-Galaxies
semi-analytic model with the same parameters as described in
\citet{LuB07}.  Energy is injected into the ICM from both SNe and AGN
following the prescription of \citet{SHT09}.  The procedure is
described in full in \citet{STY10}

By coupling a SAM to a hydrodynamical simulation, \citet{SHT09} showed
that their hybrid feedback model could reproduce the observed mean
$L_X$-$T_X$ relation for groups and poor clusters at $z=0$, but only if there was a
large energy input into the ICM from AGN over the entire formation
history of halos. The AGN heating is efficient at driving X-ray
emitting gas from the central regions of low-mass halos, reducing
their luminosity and steepening the $L_X$-$T_X$ relation as desired. Unlike
the simple preheating scenario, their model was also able to account
for some of the scatter about the mean relation seen for
temperatures $T\lesssim 3$ keV, attributable to the varied merger
histories of groups. In addition, the gas fractions of their simulated
groups and poor clusters were found to broadly agree with
observational data, rapidly declining at low temperatures and
exhibiting a comparable amount of scatter. 

The main limitation of both the PC and the FO simulations is that
neither can reproduce the low entropy found in the centres of
cool-core clusters.  For the FO run this is because cooling is not
incorporated in the hydrodynamical simulations; whereas in the PC run
preheating expels gas from cluster cores at high redshift, limiting
the subsequent cooling.

The cosmological model adopted in all three Millennium Gas simulations
is a spatially-flat $\Lambda$CDM model with parameters $\Omega_{\rm
  m,0}=0.25$, $\Omega_{\rm b,0}=0.045$, $\Omega_{\Lambda,0}=0.75$,
$h=0.73$, $n_{\rm s}=1$ and $\sigma_{8,0}=0.9$. Here
$\Omega_\mathrm{m,0}$, $\Omega_{\rm b,0}$ and $\Omega_{\Lambda,0}$ are
the total matter, baryon and dark energy density parameters,
respectively, $h$ is the Hubble parameter in units of $100$ km
s$^{-1}$ Mpc$^{-1}$, $n_\mathrm{s}$ is the spectral index of
primordial density perturbations, and $\sigma_{8,0}$ is the rms linear
density fluctuation within a sphere of radius $8h^{-1}$ Mpc. The
subscript $0$ signifies the value of a quantity at the present
day. These cosmological parameters are the same as those used in the
original Millennium simulation and are consistent with a combined
analysis of the first-year \emph{Wilkinson Microwave Anisotropy Probe}
(WMAP) data \citep{SVP03} and data from the \emph{Two-degree-Field
  Galaxy Redshift Survey} \citep{CDM01}. However, there is some
tension between the chosen parameter values, particularly $n_{\rm s}$
and $\sigma_{8,0}$, and those derived from the seven-year WMAP data
\citep{KSD10}.  More significantly for this paper, the mean
  value of the baryon density, $f_\mathrm{b}=0.18$, is higher than the
  WMAP 7-year value of $f_\mathrm{b}=0.167$.

\subsection{Cluster catalogues}

Cluster catalogues are generated at several redshifts for the three
Millennium Gas simulations using a procedure similar to that employed
by \citet{MTK02}.  Essentially, a friend-of-friends algorthm is
used to identify peaks in the density field and then spheres are grown
around these peaks until they enclose regions of a given overdensity.

We define overdensity, $\Delta$, with respect to the critical density,
$\rho_\mathrm{c}$, at any given redshift,
\begin{equation}
\Delta={\bar\rho(<r)\over\rho_\mathrm{c}}={2GM(<r)\over r^3H^2},
\end{equation}
where $\bar\rho$ is the mean density within radius $r$, $M$ is the
mass within this region and $G$ is the gravitational constant.  
We express our masses in units of $h^{-1}$\Msun, 
rescaling observational data to this system when required.

Low-mass clusters are more affected by non-gravitational heating
processes than high-mass ones.  Observationally, temperature is often
used as a proxy for mass as the two are strongly correlated.  For that
reason, we often divide the clusters into bins according to their
spectroscopic-like temperature, $T_\mathrm{sl}$, defined as
\begin{equation}
T_\mathrm{sl}={\int\rho^2T^{1\over4}\,\mathrm{d}V
\over\int\rho^2T^{-{3\over4}}\,\mathrm{d}V},
\end{equation}
where $\rho$ is the gas density, $T$ the physical temperature, and the
integral runs over volume.   \citet{MRM04} have shown $T_\mathrm{sl}$ to
be a good approximation to the temperature recovered by X-ray spectral
analysis software in the bremsstrahlung regime.

The scaling relations for the GO and PC runs can be dominated
  by small objects.  For this reason, we remove many clusters from our
  sample such that the remaining clusters are distributed evenly in
  $\log(M_{200})$, with a lower mass-limit
  $M_{2500}>1.73\times10^{13}\,h^{-1}\,\Msun$ that corresponds to 1000
  particles each of gas and dark matter within $R_{2500}$.  For the
  \Feedback\ run we resimulate all clusters with
  $kT_\mathrm{sl}>3$\,keV and a selection of clusters below this
  temperature, again chosen evenly in $\log(M_{200})$. This has a much
  higher mass resolution and so the lower mass limit in this
  case, $M_{200}>1.2\times10^{13}\,h^{-1}\,\Msun$, was instead fixed by
  the total number of clusters that we wish to simulate.  When
  plotting as a function of mass at an overdensity other than the one
  used in the selection procedure, there will not be a clean lower
  mass-limit; however this is a very minor effect that does not lead
  to any significant bias in our results.

\section{Results}
\label{sec:results}

In this section we first discuss the radial profiles of the hot gas
fractions of clusters at the present day.  We then characterise the
dependence of the hot gas fraction upon cluster mass, and investigate
the scatter about that mean relation.  Finally, we look at the
variation of the hot gas fraction with redshift.

\subsection{Profiles}
\label{sec:fgprof}

\subsubsection{Differential hot gas profiles}

\begin{figure}
\includegraphics[width=85mm]{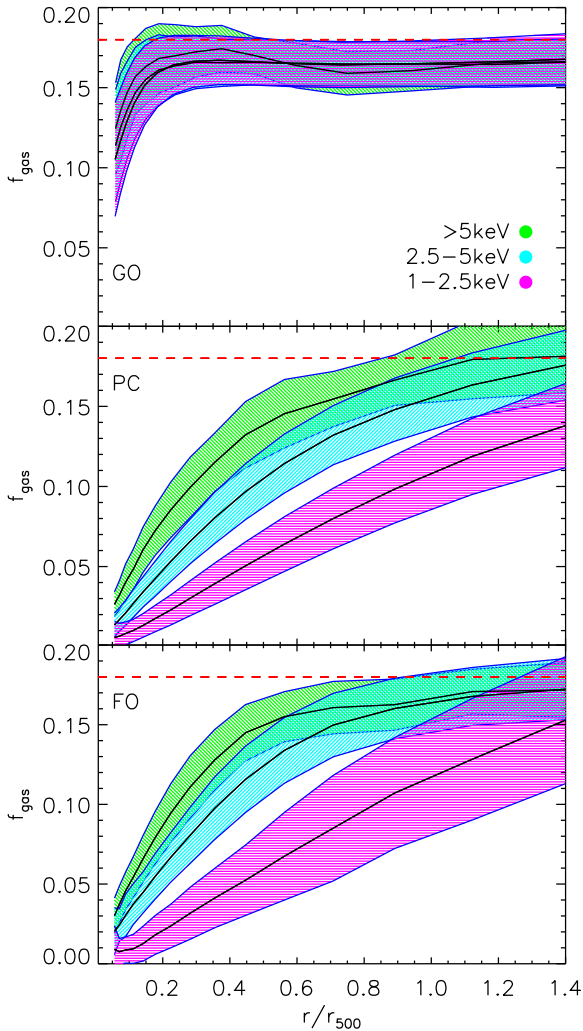}
\caption{Differential hot gas fraction profiles.  The solid lines show
  the mean relations and the shaded region the 1-sigma scatter.  The
  upper, green regions correspond to clusters with spectroscopic-like
  temperatures above 5\,keV; the middle, cyan regions to 2.5$-$5\,keV
  and the lower, magenta regions to clusters in the range
  1$-$2.5\,keV. Profiles are only plotted for radii greater than the gravitational softening length. The dashed line in each case shows the cosmic mean. }
\label{fig:fgdiffprof}
\end{figure}

Figure~\ref{fig:fgdiffprof} shows the differential gas mass fraction
profiles for clusters in three temperature ranges,
$kT_\mathrm{sl}>5$\,keV (upper, green regions),
2.5\,keV\,$<kT_\mathrm{sl}<$\,5\,keV (middle, cyan regions) and
$kT_\mathrm{sl}<$2.5\,keV (lower, magneta regions).  The curves are
plotted out to beyond $r_{500}$ (the radius at which $\Delta=500$),
which is the region accessible to X-ray observations.

For the \Nocool\ run the gas fraction plateaus at a value of 0.16--0.17 at
about $0.3\,r_{500}$, although there is a very slow increase at larger
radii (beyond the right-hand edge of the plot).  This is less than the
global baryon fraction of 0.18: conversion of kinetic energy into
heat, together with continual stirring of the gas by the motion of
dark matter structures, allows the gas to pick up energy at the
expense of the dark matter \citep[e.g.\ ][]{PTC94}.  This is
particularly evident in the cluster cores: for the largest clusters,
$r_{500}$ is of order 1\,$h^{-1}$Mpc; therefore the drop in baryon
fraction in the core of the clusters occurs on a scale significantly
larger than the force softening (25\,$h^{-1}$kpc).

In the \Precool\ and \Feedback\ runs, gas has been expelled from the
cluster cores and pushed to larger radii.  The effect is more
pronounced at lower masses: thus the gas profiles of
clusters with $kT_\mathrm{sl}<2.5$\,keV are still steeply
rising at $r_{500}$, while for $kT_\mathrm{sl}>5$\,keV the profiles
are approximately constant beyond this radius.  The inconstancy of
the gas fraction is both a nuisance, requiring careful calibration
before we can use clusters as cosmological probes, and a useful test
of any model of entropy generation in the ICM.

\subsubsection{Comparison with observations}

\begin{figure}
\includegraphics[width=85mm]{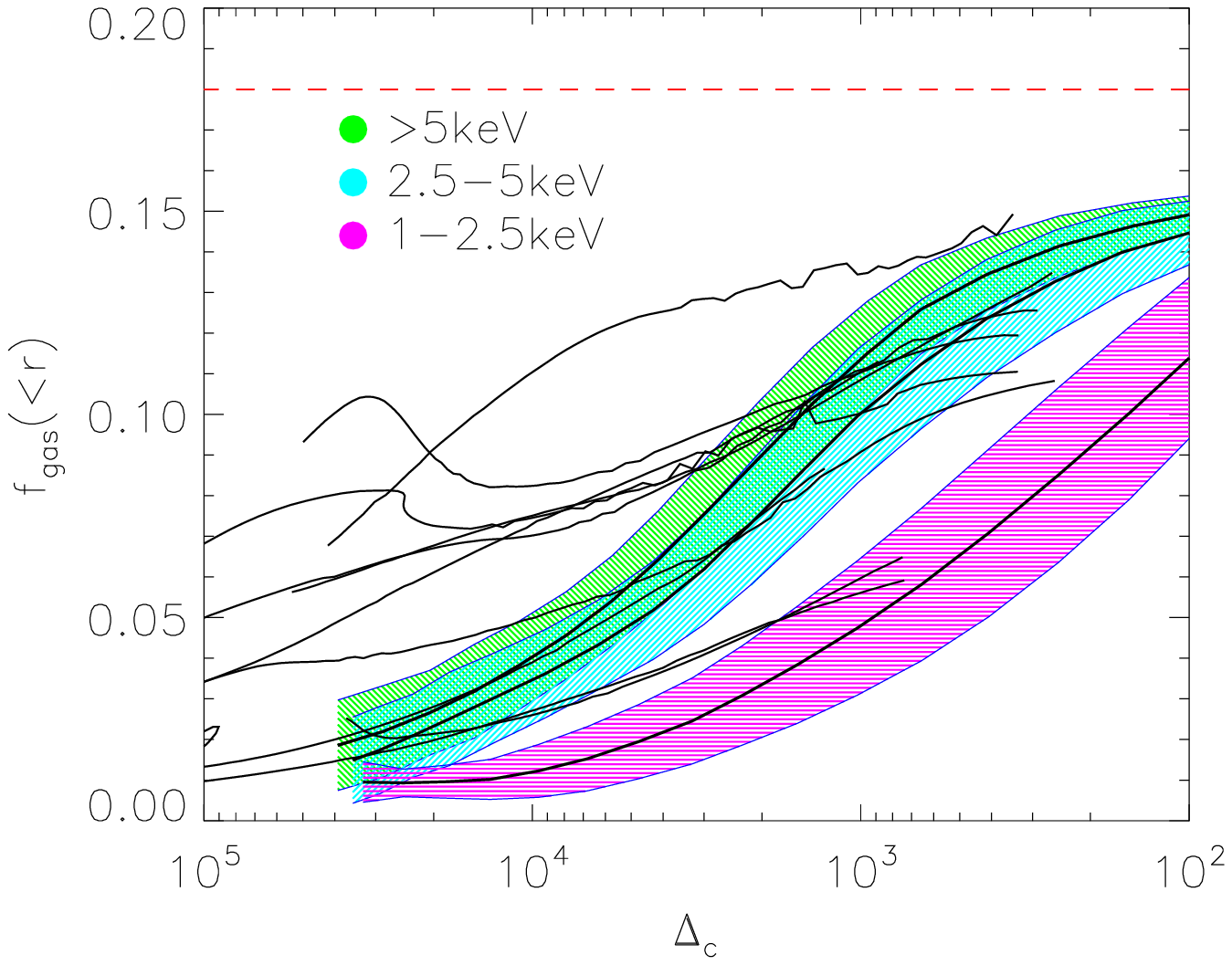}
\includegraphics[width=85mm]{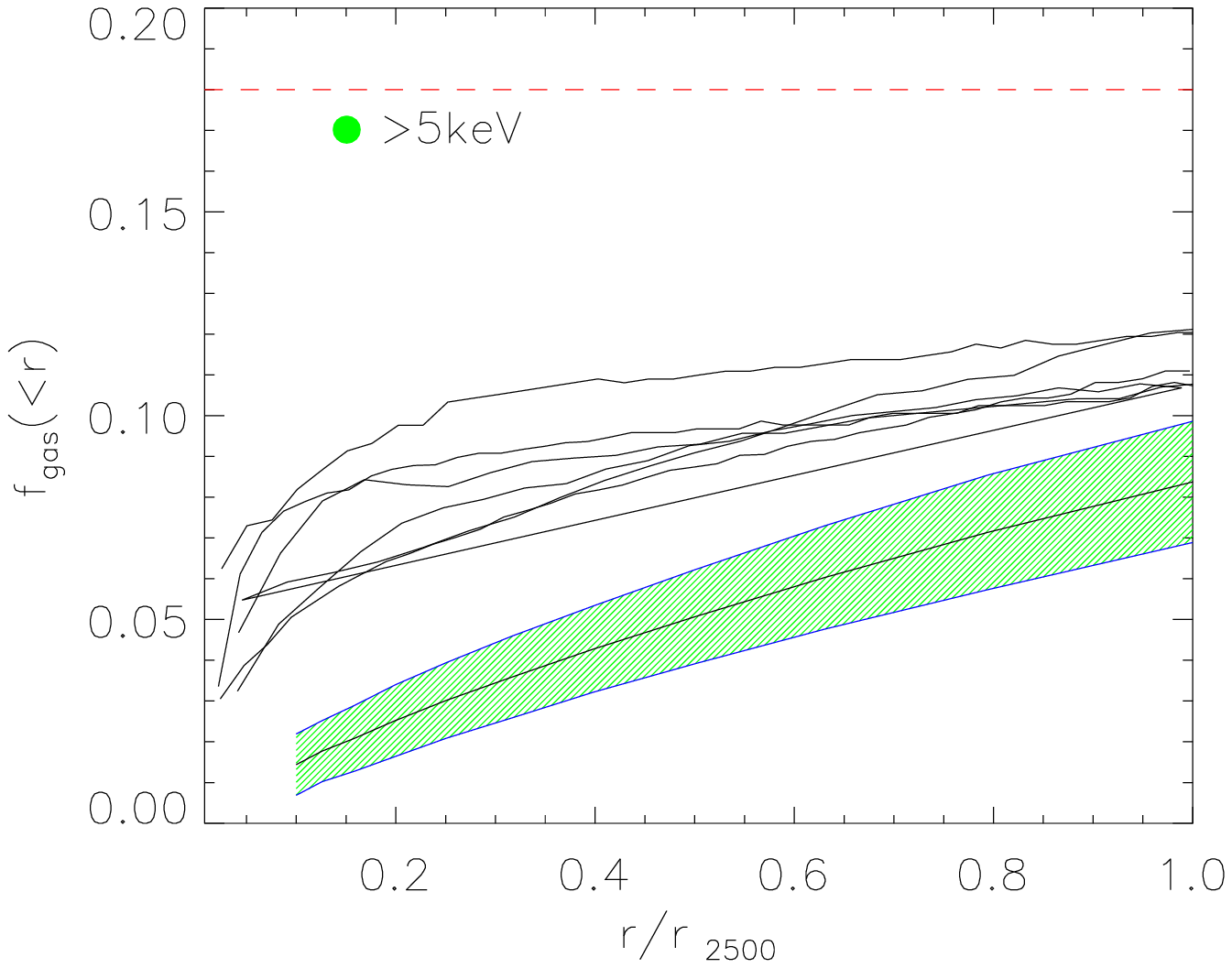}
\includegraphics[width=85mm]{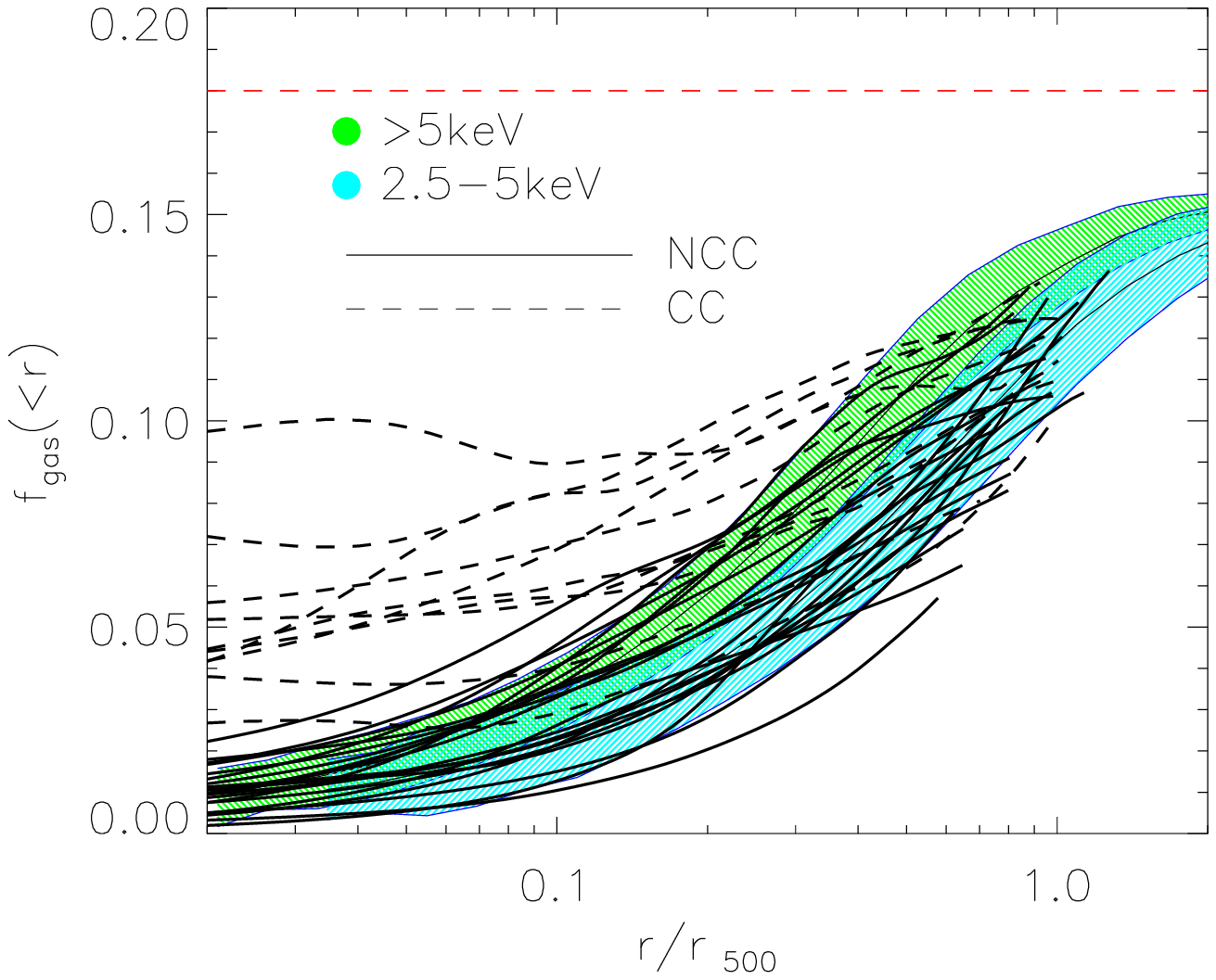}
\caption{Comparison of the cumulative hot gas fraction profiles in the
  \Feedback\ run with those of \citet[upper panel]{VKF06},
  \citet[middle panel]{ARS08} and \citet[lower panel]{PAP09}.  In each
  case the coloured bands refer to the 1-sigma spread of profiles seen
  in the simulations: upper, green ($kT_\mathrm{sl}>5$\,keV);
  middle, cyan (2.5\,keV\,$<kT_\mathrm{sl}<$\,5\,keV); lower, magenta
  (1\,keV\,$<kT_\mathrm{sl}<2.5$\,keV). The profiles of simulated clusters are only plotted for radii greater than the softening length. The observational
  data are shown by black lines with, in the lower panel, solid and
  dashed lines corresponding to non-cool-core and cool-core clusters,
  respectively.}
\label{fig:fgcumprof}
\end{figure}

Rather than plotting differential profiles, observational studies tend
to report the cumulative gas fractions, averaged within some radius.
Figure~\ref{fig:fgcumprof} shows the cumulative gas fraction profiles
in the \Feedback\ run compared to several different observational
studies \citep{VKF06,ARS08,PAP09}.  We have only plotted simulated
clusters in temperature ranges corresponding to those of the various
observational data sets.  To save space, we do not show the results
from the \Precool\ run---these are similar.

All three observational studies plot the gas fractions using a
different ordinate.  The top panel, from \citet{VKF06}, uses
overdensity relative to the critical density.  They looked at 13 {\it
  Chandra}\ clusters with a range of temperatures upwards of about
1.5\,keV for which the data extend out to large radii.  The simulated
and observed clusters agree at the outer limit of the data, but the
former fall more rapidly as one moves into the cluster
centre.

The middle panel shows data from \citet{ARS08}.  They again study {\it
  Chandra}\ clusters, but they focus on the inner regions of 42 hot
($kT_\mathrm{sl}>5$\,keV) systems.  As can clearly be seen, the
simulated clusters lie well below the observations within $r_{2500}$.

Both the above studies focus on bright, relaxed systems (those that
are likely to be labelled cool core, or CC).  By way of contrast,
the REXCESS survey \citep{BSP07,PAP09}, shown in the lower panel of
Figure~\ref{fig:fgcumprof}, is sample of 33 nearby galaxy clusters
from {\it XMM-Newton}, selected so as to sample a broad range of
luminosities and with no bias towards any morphological type.  The
temperature range here is 2-9\,keV with the more massive clusters
lying towards the upper edge of the observed band, and the least
massive ones towards the bottom.  Here the simulations are much more
successful in reproducing the observed profiles, providing a fair
match to the non-cool-core (NCC) population out to the limit of the
observations.  They fail to reproduce CC clusters (those with
flattened baryon fraction profiles in their inner regions), however
these are much less frequent than in the relaxed samples.

We conclude that our simulated clusters provide a fair match to the
hot gas fractions in typical NCC clusters, but fail to
reproduce the higher gas fractions seen in the central regions of
the brighter, CC clusters.


The results for our adiabatic halos agree with other previous
simulations undertaken with SPH \citep[e.g.~][]{KNV05,EDB06,CEF07}.
Direct comparison of the other runs is more difficult because we use
different feedback models.  

Qualitatively, we see a similar behaviour in the profiles of the hot
gas fraction to previous work, but the baryon fraction profile is very
different in our \Feedback\ run because of the much reduced stellar
fraction.  The integrated baryon fractions within $r_{500}$
are considered in Section~\ref{sec:scaling:baryons}.

\subsection{Scaling relations}
\label{sec:scaling}

\subsubsection{Cumulative hot gas fractions}
\label{sec:scaling:hotgas}

\begin{figure}
\includegraphics[width=85mm]{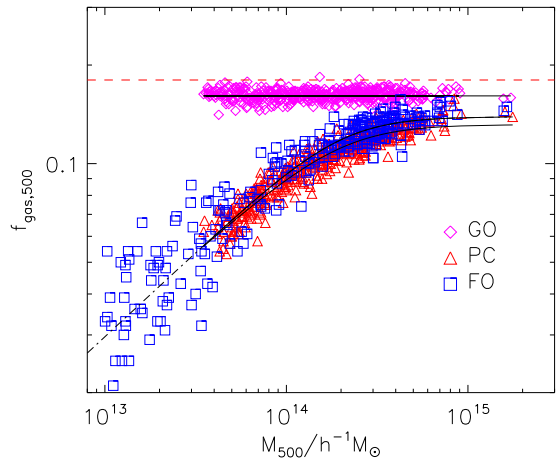}
\caption{The cumulative hot gas fraction within $r_{500}$ as a 
function of total mass.  The solid
  lines show the best-fitting mean relations, as described in the text
  and Table~\ref{tab:fgfit}.  Extending the fit to the whole range of
  the \Feedback\ data, as shown by the dot-dashed line, makes almost
  no difference to the fit.  The dashed line shows the mean baryon
  fraction in the simulation. }
\label{fig:fgcuma}
\end{figure}

Figure~\ref{fig:fgcuma} shows the cumulative gas fraction within a
radius of $r_{500}$ as a function of total mass.  
The \Nocool\ points are consistent with a
constant value of 0.162, slightly smaller than the universal mean of
0.18.  By contrast, both the \Precool\ and the \Feedback\ runs have
hot gas fractions that are strong functions of mass, because the
feedback processes are more effective in lower-mass clusters and
evacuate more of the gas.

\begin{figure}
\includegraphics[width=85mm]{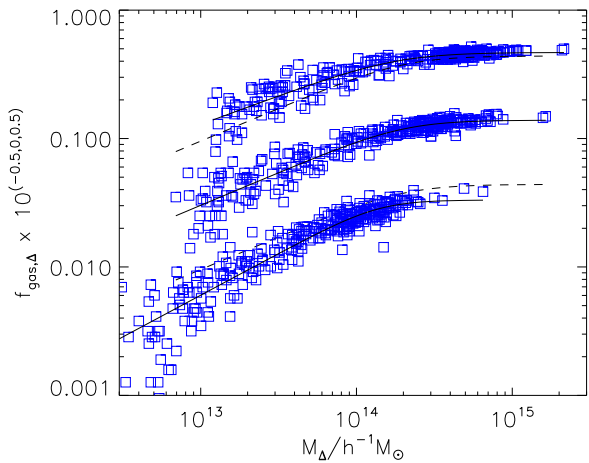}
\caption{The baryon fraction within three different radii for the
  \Feedback\ run: $r_{200}$ (upper), $r_{500}$ (middle) and $r_{2500}$
  (lower).  For clarity, the upper and lower data points have been
  shifted by 0.5 dex.  The solid lines show the best-fitting mean
  relations and the dashed lines the best-fitting mean relation for
  the $r_{500}$ data.}
\label{fig:fgcumb}
\end{figure}

Figure~\ref{fig:fgcumb} contrasts the hot gas fractions for the
\Feedback\ run within three different radii corresponding to enclosed
overdensities of $\Delta=200$, $500$, and $2500$ (the \Precool\ run gives
similar results).  In each case the mass has been measured within the
corresponding radius.  As expected from the radial profiles, the gas
fraction is an increasing function of scale radius (i.e.~decreasing
overdensity).  Note that, for a fixed enclosed mass, the variation in
enclosed gas fraction is relatively modest: e.g.~at
$M_\Delta=10^{14}\,h^{-1}$\,\Msun\ it varies from 0.07 for
$\Delta=2500$ to 0.10 for $\Delta=200$.  This is much less than the
variation seen if a fixed overdensity is used to measure the mass,
e.g.~for $M_{500}=10^{14}\,h^{-1}$\,\Msun, the enclosed gas fraction
increases from 0.04 to 0.11 as the overdensity drops from 2500 to 200.

In Figures~\ref{fig:fgcuma} and \ref{fig:fgcumb}, we fit the hot gas
fraction scaling relations with models of the form
\begin{equation}
\log_{10}f=\log_{10}f_0+s\big(\mu-\log\big(1+\exp(\zeta\mu))/\zeta\big),
\label{eq:fgmodel}
\end{equation}
where $\mu=\log_{10}(M/M_0)$ and $f_0$, $M_0$, $\zeta$ and
$s$ are fitting parameters.  In log space, this represents a line of
constant slope, $s$, at masses well below $M_0$, bending over to a
constant value of $f_0$ at high masses.  For low enclosed
overdensities we would expect $f_0$ to tend towards the universal
baryon fraction of 0.18, although we do not impose this as a
constraint.  $\zeta$ is a parameter that controls the abruptness of
the transition between the two regimes.  The data are not
  always sufficient to independently constrain all the parameters, and
  in particular $\zeta$: for that reason we use a fixed value of
  $\zeta$ when recording our fits.  The best-fit models are shown as
solid lines in the figures, and the parameters are listed in
Table~\ref{tab:fgfit}, along with the scatter about the best-fit
relation.

\begin{table}
\begin{center}
  \caption{Model parameter fits to the hot gas fractions as a function
    of mass, as described by Equation~\ref{eq:fgmodel} with a
      fixed value of $\zeta=4$.  Here $m=\log_{10}(M_0)$.  Typical
    1-sigma errors in $f_0$, $m$ and $s$ are 0.005, 0.2 and 0.04,
    respectively.  For the \Nocool\ run there is no discernible
    mass-dependence in the hot gas fractions and so only the mean
    value is recorded in column $f_0$.  The final column gives the
    root-mean-square scatter in dex of the data points about the
    best-fit line.}
\label{tab:fgfit}
\begin{tabular}{rrcccc}
\hline
Model& Overdensity& $f_0$& $m$& $s$& $\sigma$\\
\hline
\Nocool& 2500& 0.158& & &0.036\\
      &  500& 0.161& & &0.018\\
      &  200& 0.163& & &0.013 \\
      & 500$\backslash$2500& 0.164& & &0.026 \\
      & 200$\backslash$500& 0.166& & &0.033 \\
\Precool& 2500& 0.103& 14.15& 0.590& 0.048\\
      &  500& 0.134& 14.26& 0.519& 0.041\\
      &  200& 0.150& 14.43& 0.363& 0.027\\
      & 500$\backslash$2500& 0.161& 14.28& 0.393& 0.037\\
      & 200$\backslash$500& 0.180& 14.10& 0.263& 0.044\\
\Feedback& 2500& 0.126& 14.25& 0.650& 0.124\\
      &  500& 0.143& 14.26& 0.552& 0.061\\
      &  200& 0.148& 14.21& 0.472& 0.058 \\
      & 500$\backslash$2500& 0.163& 14.13& 0.492& 0.076\\
      & 200$\backslash$500& 0.173& 13.72& 0.512& 0.066\\ 
\hline
\end{tabular}
\end{center}
\end{table}

\begin{figure}
\includegraphics[width=85mm]{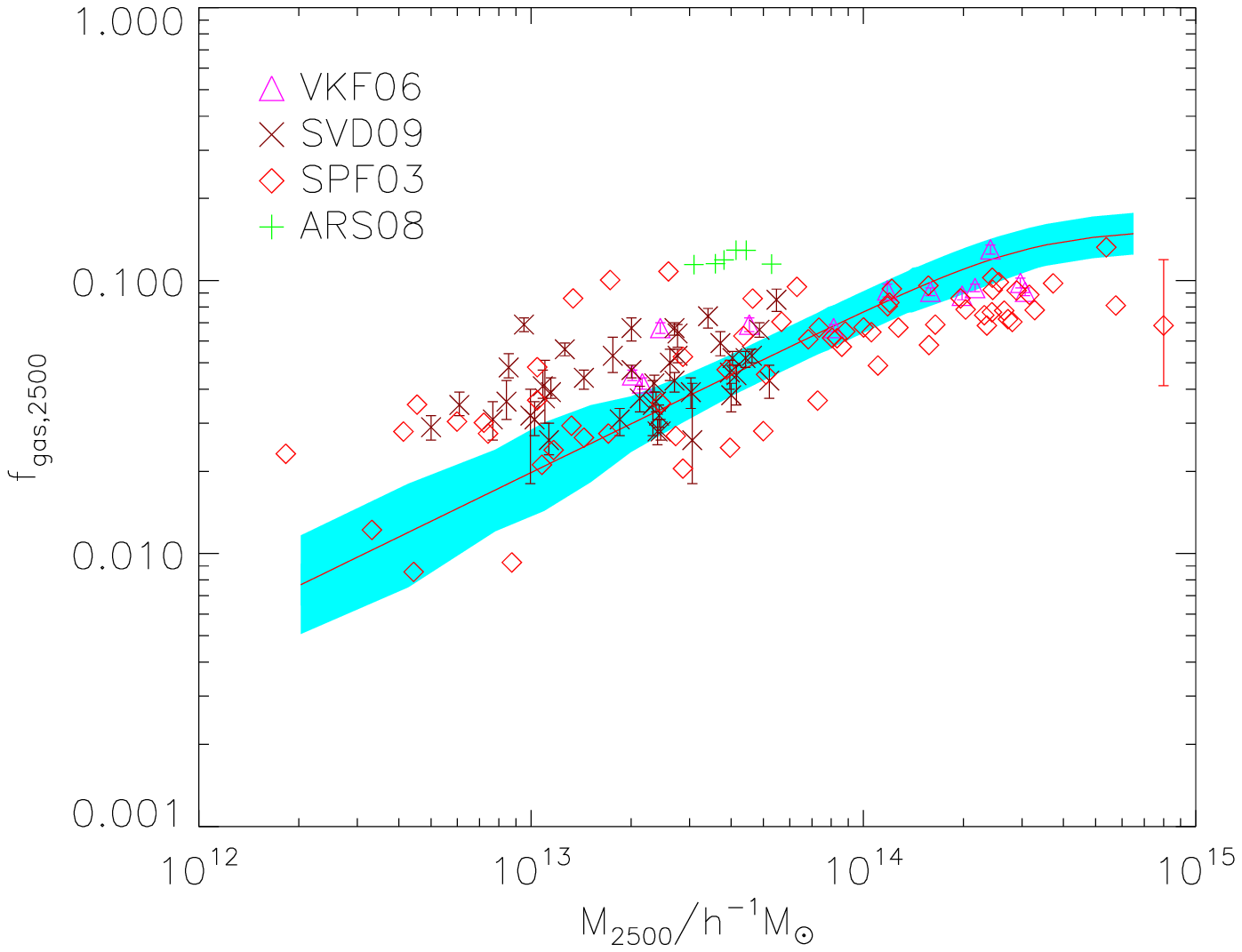}
\includegraphics[width=85mm]{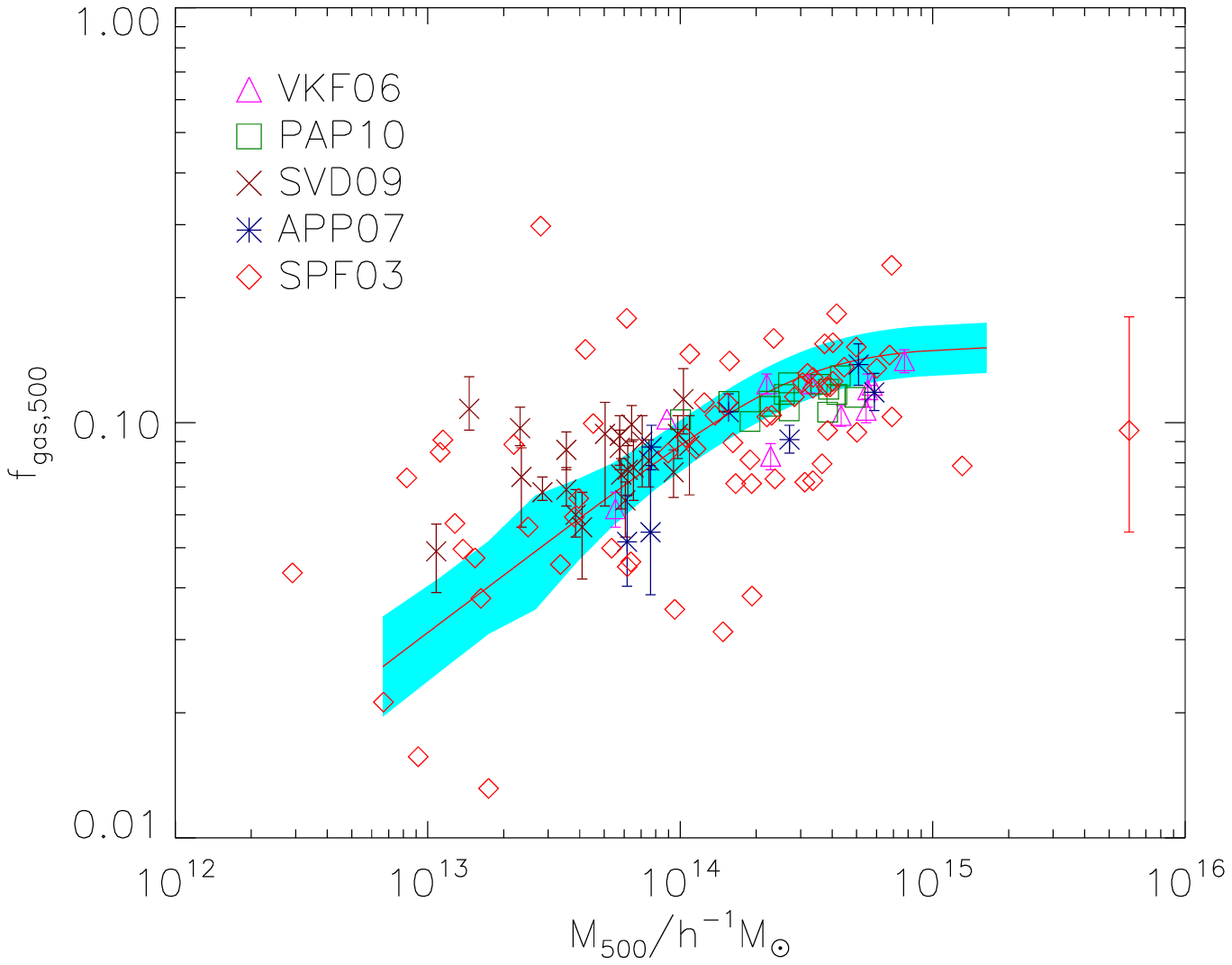}
\caption{The cumulative hot gas fraction versus mass relations for the
  \Feedback\ run as compared to observations.  The upper and lower
  panels refer to overdensities of 2500 and 500, respectively.  The
  shaded regions are the 1-sigma spread in the simulated clusters.
  The symbols represent observational data from
  \citet{SPF03}, \citet{VKF06}, \citet{APP07}, \citet{ARS08}, \citet{SVD09} and \citet{PAP09}.  For the \citet{SPF03}
  clusters we show one fake data point on the right-hand edge of the
  plots with typical 1-sigma (statistical) error bars. For
  the other samples, we include the error bars on the plotted points,
 apart from \citet{ARS08} and \citet{PAP09} since we do not know the uncertainties in 
their measurements.}
\label{fig:fgobs}
\end{figure}

Figure~\ref{fig:fgobs} shows the cumulative hot gas fractions in the
\Feedback\ run within $r_{2500}$ and $r_{500}$ as a function of mass.
The shaded region shows the 1-sigma spread about the mean relation from
the simulation and the points are observational data from the sources
listed in the figure caption.  Concentrating first on the upper panel,
it is apparent that there are systematic differences in the
  reported hot-gas fractions that cannot be attributed solely to
  statistical error.  In particular, the \citet{ARS08} data lie
  significantly above those of the other studies.  This may be because
  they concentrated on regular, CC clusters that are likely to lie up
  the upper edge of the distribution.  Even given these observational
  inconsistencies, however, the simulations show a much greater
  variation in gas fraction with $M_{2500}$ than do the observations.
It would seem that in the \Feedback\ run (and the \Precool\ run is
similar) we have ejected too much material from within this radius in
small clusters, and too little in large ones.

Moving out to $r_{500}$, as shown in the lower panel of the figure,
the simulations and observations are in better agreement.  There
is again a suggestion that the observational data would prefer higher
gas fractions than the simulations below a cluster mass of
of $5\times10^{13}h^{-1}$\Msun, but the scatter in the observational
measurements is large.  At higher masses, the two show a similar trend
of increasing gas fraction with cluster mass, although the simulated
values are perhaps slightly too high. This is presumably because the mean baryon
fraction that we have used in the simulations, 0.18, is higher than
the current WMAP best fit value of 0.168.

\subsubsection{Differential hot gas fractions}

The profiles of Figure~\ref{fig:fgdiffprof} suggest that the
differential gas fraction between radii of $r_{2500}$ and $r_{500}$
may provide a measure that is more independent of mass than the
cumulative gas fractions of the previous section.

\begin{figure}
\includegraphics[width=85mm]{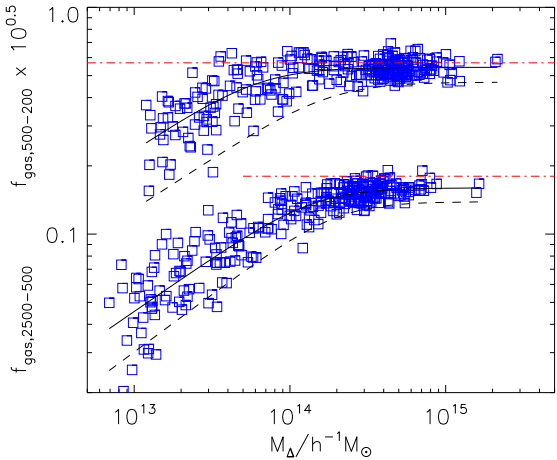}
\caption{Differential hot gas fractions for the \Feedback\ runs:
  between $r_{2500}$ and $r_{500}$ (lower points), and between
  $r_{500}$ and $r_{200}$ (upper points).  The
  upper points have been shifted up by 0.5 dex, for clarity.  In each
  case, the mass has been taken to be that at the outer edge of the
  differential range.  The solid lines show the best-fit mean relation
  and the dashed lines show the best-fit relation for the equivalent
  cumulative gas fraction measure. The dash-dotted lines show the mean
  baryon fraction in the simulation.}
\label{fig:fgdiff}
\end{figure}

\begin{figure}
\includegraphics[width=85mm]{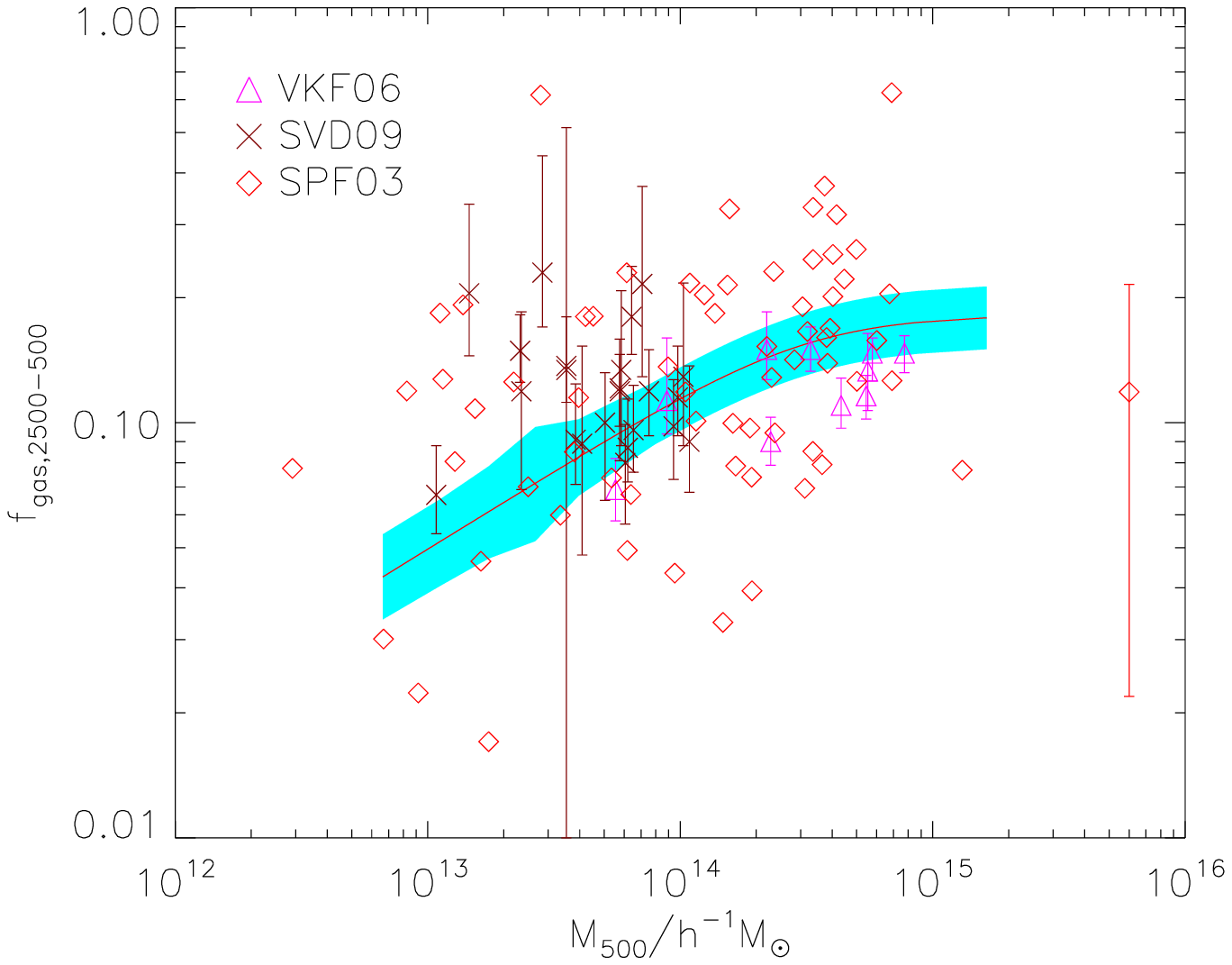}
\caption{The differential hot gas fraction in the annulus contained
  between $r_{2500}$ and $r_{500}$ for the \Feedback\ run as compared
  to observations.  The shaded region is the 1-sigma spread in the
  simulated clusters.  The points are taken from
  \citet{SPF03}, \citet{VKF06} and \citet{SVD09}.  For the \citet{SPF03}
  clusters we show one fake data point on the right-hand edge of the
  plot with typical 1-sigma (statistical) error bars. For
  the other samples, we draw the error bars on the plotted points.}
\label{fig:fgdiffobs}
\end{figure}

In Figure~\ref{fig:fgdiff} we show differential gas fractions,
$r_{2500}$-$r_{500}$ and $r_{500}$-$r_{200}$, for the \Feedback\ model
(once again, the \Precool\ run gives similar results).  Both are
higher than the equivalent cumulative measures, although the universal
gas fraction is reached only for the most massive clusters
($M_{200}>3\times10^{14}\,h^{-1}\,\Msun$) at radii $r>r_{500}$.
Clearly the differential gas fraction at larger radii, $r_{500}$--$r_{200}$, is more
nearly constant and so provides the more accurate probe of cosmology,
but observationally the inner annulus, $r_{2500}$--$r_{500}$, provides
a compromise between eliminating the depleted inner region and having
enough counts to enable a reliable X-ray determination of the gas
density.  The observational data from \citet{SPF03}, \citet{VKF06} and
\citet{SVD09}, plotted in Figure~\ref{fig:fgdiffobs},
present a confused picture.  The \citet{SPF03} data broadly mimic the
  simulations, but the statistical scatter in their data is very
  large.  \citet{VKF06} report the smallest error bars for their data
  and find differential gas fractions that increase strongly with
  mass, but which lie below the simulated values; whereas, at
  lower masses, the \citet{SVD09} data seem to require hot gas
  fractions that are decreasing, or at best flat, as a function of
  mass.  We conclude that these differential measurements are not yet
  sufficiently robust to provide useful constraints.

It is interesting to note that both the \Precool\ and
\Feedback\ models have higher differential gas fractions at large
radii than does the \Nocool\ model.  The injection of entropy has
removed gas from the cores of the clusters and pushed it out to larger
radii, between $r_{500}$ and $r_{200}$. In a steady-state, the higher entropy in these runs would ensure that they have a lower gas density than in the GO model: we conclude that on large scales, although still within the virial radius, the clusters are not in dynamical equilibrium.

\subsubsection{Baryon fractions}
\label{sec:scaling:baryons}

The baryon fractions are also well-fit by the model given in
Equation~\ref{eq:fgmodel} with parameters as listed in
Table~\ref{tab:fbfit}.  Because clusters are large systems with deep
potential wells, it is often stated that they should enclose a
representative sample of the Universe, and in particular that they
should contain the Universal fraction of baryons.  Indeed, we find
this is approximately true, with only a small baryon deficit within
$r_{200}$.  

\begin{table}
\begin{center}
  \caption{As for Table~\ref{tab:fgfit} but for the baryon fractions
    rather than the hot gas fractions. $\zeta=$ is fixed at 4
    for the \Precool\ and 8 for the \Feedback\ run.
    For the \Nocool\ run there is no star formation and so the values
    are the same as reported in Table~\ref{tab:fgfit}.}
\label{tab:fbfit}
\begin{tabular}{rrcccc}
\hline
Model& Overdensity& $f_0$& $m$& $s$& $\sigma$\\
\hline
\Precool& 2500& 0.141& 14.32& 0.250& 0.061\\
      &  500& 0.161& 14.44& 0.307& 0.037\\
      &  200& 0.168& 14.49& 0.271& 0.030\\
      & 500$\backslash$2500& 0.174& 14.29& 0.362& 0.035\\
      & 200$\backslash$500& 0.188& 14.10& 0.269& 0.042\\
\Feedback& 2500& 0.156& 14.92& 0.251& 0.069\\
      &  500& 0.150& 14.48& 0.308& 0.052\\
      &  200& 0.157& 14.41& 0.291& 0.044 \\
      & 500$\backslash$2500& 0.167& 14.25& 0.357& 0.053\\
      & 200$\backslash$500& 0.181& 13.94& 0.333& 0.058\\ 
\hline
\end{tabular}
\end{center}
\end{table}

Observationally, the baryon fraction is hard to determine because of
the difficulty in measuring the contribution from dwarf galaxies and
from intracluster light (stars that have been stripped from galaxies).
This latter component may comprise as much as 40 percent of the total
light of the cluster \citep{BNT95,GZZ00,FMM02,FMM04,GZZ05,KBP06,ZWS05}.
Except for the brightest X-ray clusters, the measurement of total mass
is also problematic.

\begin{figure}
\includegraphics[width=85mm]{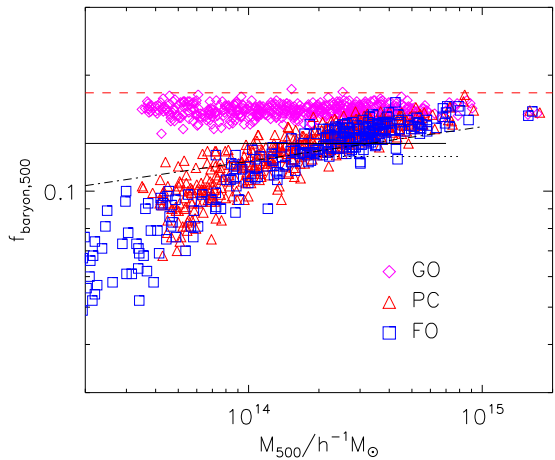}
\caption{The cumulative baryon fraction versus mass.  The solid and
  dotted lines show the mean relations (approximately independent of
  mass) from \citet{GZZ07} and \citet{LLA08}, respectively; the
  dash-dotted line shows that from \citet{GPF09}.  The red dashed line
  is the universal mean.}
\label{fig:fb}
\end{figure}

\citet[hereafter GZZ07]{GZZ07}, in a sample of 12 groups and clusters
spanning a wide mass range, find that the baryon fraction is
independent of mass and averages to 0.133 within $r_{500}$.
\citet[hereafter LLA08]{LLA08} with a smaller sample of 5 high-mass
clusters find a slightly lower value of 0.123 (the mean of their
quoted numbers, weighted by the inverse square of their errors).  On
the other hand \citet[hereafter GPF09]{GPF09}, in a sample of 41
clusters drawn from \citet{VKF06}, \citet{APP07} and \citet{SVD09},
find that the baryon fraction is a slowly-increasing function of mass.
The baryon fractions for our \Feedback\ and \Precool\ clusters are
compared to these observations in Figure~\ref{fig:fb}

The observations and the simulations approximately agree for cluster
masses of 1--$3\times10^{14}h^{-1}\,$\Msun. However the simulations
show a strong variation with cluster mass, even more so than that of
GPF09.  This difference is attributable mainly to differences in
star formation, as described below.

\subsubsection{Stellar fractions}
\label{sec:scaling:stars}

\begin{figure}
\includegraphics[width=85mm]{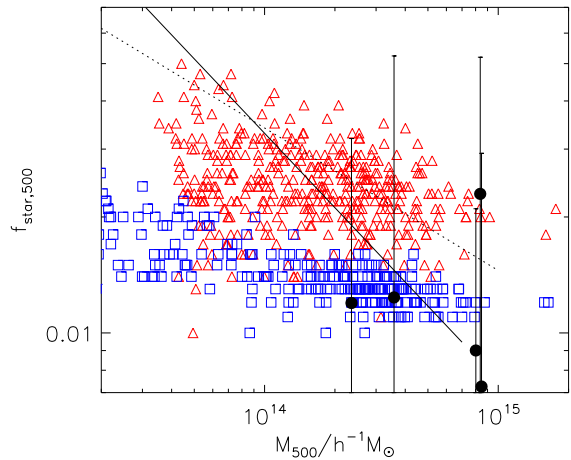}
\includegraphics[width=85mm]{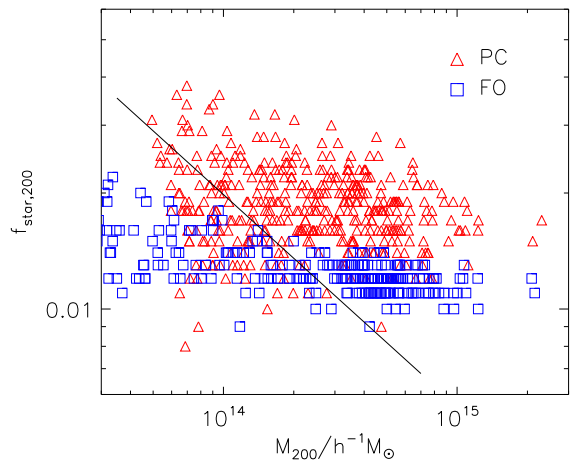}
\caption{The cumulative stellar fraction versus mass.  The upper panel
  shows the stellar fractions within $r_{500}$; the black dots are
  observed clusters from LLA08, the solid line shows the observed
  relation from GZZ07, and the dotted line that from GPF09.  The lower
  panel is the stellar fraction within $r_{200}$, with the solid line
  showing the observed relation from \citet{And10}.}
\label{fig:fs}
\end{figure}

Both our \Precool\ and our \Feedback\ models
have much more modest star formation than do many previous simulations
\citep[e.g.~][]{BMS04,EDB06,KSA07,NKV07,DOS08,FBT10,PSS10}.  In particular,
the \Feedback\ run takes its star-formation rate from the
highly-successful L-Galaxies semi-analytic model.  The mean stellar
fraction in our high mass clusters in the \Feedback\ run agrees well
with the observations of both GZZ07 and LLA08; the results of GPF09
are slightly higher.  However, we do not find such a strong increase
in stellar fraction in low-mass clusters as is seen in both GZZ07 and
GPF09.  The upper panel in Figure~\ref{fig:fs} shows the stellar
mass-fraction within $r_{500}$ as a function of mass for both the
\Precool\ and the \Feedback\ runs, with the trend from GZZ07 shown as a
solid line and that from GPF09 as a dotted line.  This comparison
suggests that we considerably underestimate star formation in groups.
The lower panel shows a similar plot for $r_{200}$ with data from
\citet{And10}.  He finds lower stellar fractions but a similar steep
increase with decreasing mass.

Before dismissing our \Feedback\ model as unrealistic, however, we
note the following: 
\begin{itemize}
\item As pointed out by \citet{BMB08a}, the GZZ07 data is incompatible
  with any model that forms galaxies via hierarchical mergers unless
  there is an unreasonably large star-formation rate in groups at late
  times.
\item The L-Galaxies model produces correlation functions for the
  galaxy distribution that are consistent with observations with no
  evidence for a suppression at small separations \citep{KiW08}.  It
  is difficult to reconcile this with the need to greatly increase the
  stellar fraction in groups.
\end{itemize}

The other simulations mentioned above also predict a slow variation of
stellar fraction with mass, although they mostly have stellar
fractions that are higher than ours, agreeing with the observations on
group scales but having stellar fractions that are too high on cluster
scales.  For example, the clusters of \citet{EDB06} have a stellar
fraction of about 0.05 within $r_{500}$.  The equivalent fraction for
massive clusters ($kT_\mathrm{sl}>5$\,keV) in our own runs is 0.02 for
\Precool\ and 0.013 for \Feedback.
\footnote{\citet{EDB06} choose to give stellar fractions in terms of
the mean baryon fraction,
$Y=M_\mathrm{star}/M_\mathrm{total}/f_{\rm b}$.   The
stellar fractions quoted in their paper are thus a factor of 5.6
larger than those listed here.}

We conclude that, although the observations are not yet sufficiently
robust, stellar mass fractions provide an important test of, and
discriminant between, different galaxy formation models.

\subsubsection{Scatter in the scaling relations}
\label{sec:scatter}

In this section we investigate why some clusters have slightly more
hot gas, and others slightly less, than other clusters of the same
mass.  Our purpose in doing this is two-fold: firstly to understand
the physical reason for this scatter, and secondly to suggest
corrections that can be applied to the observations to better allow
gas fraction to be used as a probe of cosmology.

First note that, as is evident from Figure~\ref{fig:fgcuma}, the mean
gas fractions in the \Nocool\ run are independent of mass.
However, there is scatter about the mean gas fraction that we might
hope to relate somehow to the physical properties of the cluster.

\begin{figure}
\includegraphics[width=85mm]{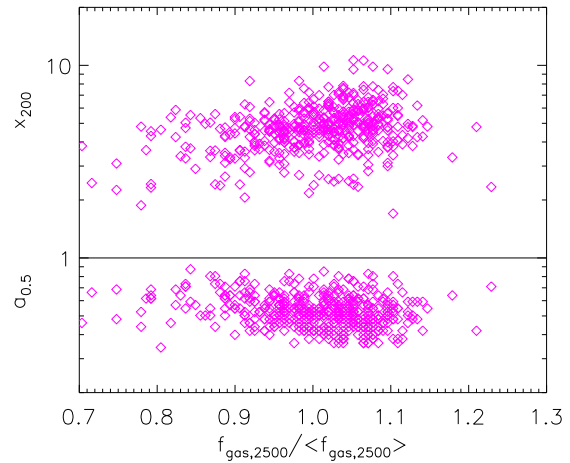}
\caption{The strongest correlations of the scatter in the mean gas
  fraction within \RE\ for the \Nocool\ run.  The upper points show the
  cluster concentration, and the lower points the expansion factor of
  the Universe at the time that the cluster had accumulated half its
  final mass.}
\label{fig:fgnocoolcorr}
\end{figure}

We have checked for correlations of the scatter with every conceivable
physical quantity (including substructure, merging history, angular
momentum, etc.) and find many weak correlations, but no strong one.
Figure~\ref{fig:fgnocoolcorr} shows a positive correlation with
concentration, i.e.~more concentrated clusters have a greater gas
fraction than the average within \RE.  Likewise, clusters that form
earlier have a greater gas fraction than those that form later.
The correlation coefficients for these two relations are 0.29 and
$-0.23$, respectively.
These may be two aspects of the same relation as concentration shows a
negative correlation with formation time.  The appendix describes how
we measure each of these for the clusters in our simulation.  Both are
correlated with cluster mass, but it turns out that they are more
strongly correlated with each other.

The physical mechanism that may drive the correlations seen in
Figure~\ref{fig:fgnocoolcorr} is unclear: it primarily affects the
core of the cluster as the correlations get weaker if one measures the
gas fraction within larger radii.  It may be that the degree of
gravitational pre-heating is greater in systems that form later
\citep[see, e.g.,][]{MYB05}.

Other quantities that we have tested include the halo angular
momentum, merger history and substructure.  Once the primary
correlation of gas fraction with mass is removed, none of these
show any correlation with the residual gas fraction.

\begin{figure}
\includegraphics[width=85mm]{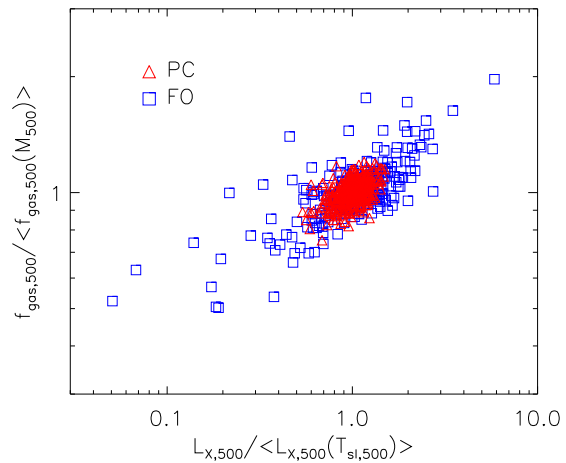}
\caption{The deviation from the mean gas fraction-mass relation
plotted against the deviation from the mean luminosity-temperature relation.
  Properties are measured within a radius of $r_{500}$.
  Shown are the ratio of the measured quantities compared to that of
  the best-fitting mean relation.}
\label{fig:fglxtsl}
\end{figure}

Observationally, of course, an excess core gas fraction is
associated with an increase in X-ray luminosity.  Thus, as shown in
Figure~\ref{fig:fglxtsl}, the excess luminosity of a cluster above the
mean $L_\mathrm{X}$-$T_\mathrm{sl}$ relation is correlated with the
presence of excess gas in the cluster.  The correlation
coefficients in this case are 0.51 and 0.73 for the \Precool\ and for
the \Feedback\ run, respectively.  A similar result was found for
the \Precool\ run by \citet{SRE09}.  This correlation may serve to
reduce the scatter from the major outliers in the gas fraction-mass
relation.

\subsection{Evolution}
\label{sec:evolution}

\subsubsection{Profiles}

\begin{figure}
\includegraphics[width=85mm]{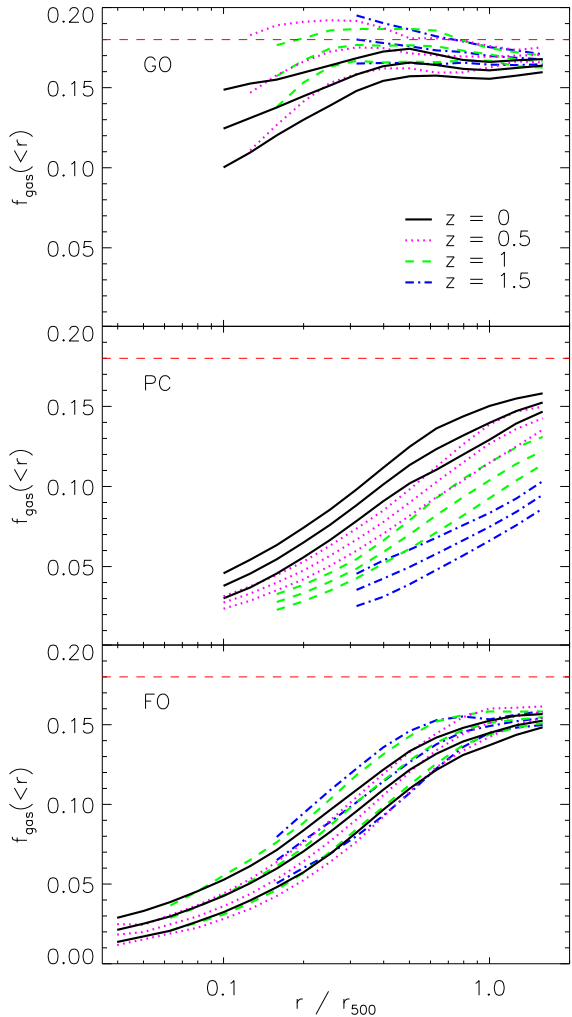}
\caption{The evolution of the cumulative gas fraction profiles of the 10
  most massive clusters.  In each case the dot-dashed blue, dashed
  green, dotted magenta and solid black lines correspond to $z=$1.5, 1, 0.5
  and 0, respectively.  The middle line of each set is the mean of,
  and the upper and lower the 1-sigma spread in, the gas
  fraction profiles. The red dashed line is the universal mean.}
\label{fig:fgevol}
\end{figure}

Figure~\ref{fig:fgevol} shows the evolution of the cumulative gas
fraction profiles of the 10 most massive clusters, for each of the runs.

Looking first at the \Nocool\ run in the upper plot, it can be seen
that the gas fraction within $r_{500}$ remains largely constant over
time: this is to be expected for self-similar evolution.  As the
effective resolution increases (i.e.~the ratio of the smoothing length
to $r_{500}$ decreases) so the gas fraction within the cluster core
can be seen to be depleted, though still much higher than in the
other two runs.

In the \Precool\ simulation, the gas fractions at high redshift are
much reduced over their current-day values.  This is because a large
amount of energy has been injected into the ICM at early times,
expelling gas from the clusters.  Subsequently, the gas falls back
into the clusters as the Universe evolves and tends towards (but falls
far short of) self-similar evolution.  In other words, the
early entropy injection becomes relatively less important in more massive
systems at late times.

This is in contrast to the behaviour in the \Feedback\ run.  Here we
have continual injection of energy so that gas fraction profiles
remain constant over time.  Although our simulation takes its level
of feedback from a semi-analytic model, nevertheless it seems to
have achieved a homologous evolution.

Thus, although the \Precool\ and \Feedback\ clusters have
indistinguishable gas profiles at the current day, they look very
different in the past.  This casts doubt on the use of the measured
gas fraction as a cosmological probe, but instead opens the
possibility that it can be used to determine the nature of the
feedback mechanism: in particular to distinguish between early
(\Precool) and continual (\Feedback) heating.



\subsubsection{Scaling relations and comparison with observations}

\begin{figure}
\includegraphics[width=85mm]{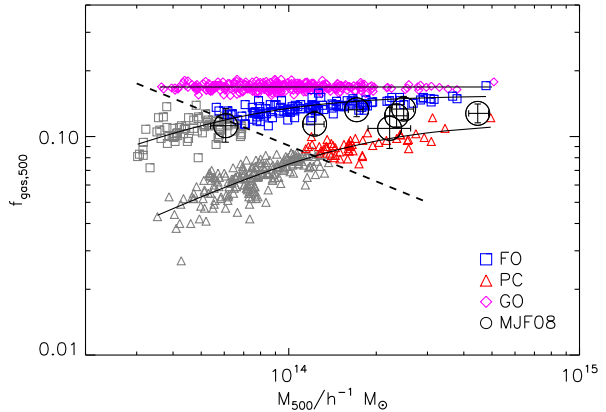}
\caption{Gas fractions within $r_{500}$ at $z=1$ as a function of 
total mass.  The solid lines
  show the best-fit mean relations. The black circles are
    high-redshift clusters ($0.8\leq z\leq 1.3$) from the
    observational dataset of \citet{MJF08}. The dashed line
    illustrates the effect of imposing a typical survey flux limit of
    $6.5\times 10^{-14}$\,erg\,s$^{-1}$\,cm$^{-2}$ at $z=1$; only clusters to the
    right of this line would actually be observed by such a survey (the observed cluster slightly to the left of this line is at a lower redshift $z\approx 0.8$).}
\label{fig:fgz1}
\end{figure}

Figure~\ref{fig:fgz1} shows the hot gas fractions within
  $r_{500}$ at a redshift $z=1$ for clusters in our three
  simulations. For comparison, we also plot high-redshift systems
  ($0.8\leq z\leq 1.3$) from a catalogue of clusters observed with
  {\it Chandra} compiled by \citet{MJF08}. Note that \citet{MJF08} do
  not themselves present $f_{\rm gas}$ values. To compute them, we
  first determine the total mass, $M_{500}$, from the supplied values
  of $Y_{\rm X}$ (where $Y_{\rm X}$ is defined as the product of the
  gas mass within $r_{500}$ and the spectroscopic-like temperature in
  the spherical annulus $0.15\,r_{500}<r\leq r_{500}$) by using the
  $Y_{\rm X}-M_{500}$ relation derived from the sample of
  \citet{VKF06}.  This is the procedure adopted by \citet[see their
    Equation 4]{MJF08}. The gas fraction then follows upon taking the
  ratio of the gas mass interior to $r_{500}$ (tabulated in their
  paper) to the total mass. The errors on the observational data
  points in Figure~\ref{fig:fgz1} are computed using the supplied
  statistical errors on the gas mass and the core-excised
  temperatures. We compute errors on $f_{\rm gas}$ in this way, rather
  than using the errors on $Y_{\rm X}$, because the gas mass and
  temperature are independent measurements.

The different evolutionary behaviour of the gas fraction
  profiles is reflected in that of the mean gas fractions: the
  \Precool\ clusters have significantly smaller gas fractions at early
  times than those in the \Feedback\ run. On the whole, the
  predictions of the FO model provide a closer match to the
  observational data than those of the PC model, but it seems as if
  some of the observed data points lie below the lower edge of the FO
  $f_{\rm gas}-M_{500}$ relation. This could be because the mean
  cosmic baryon fraction in our simulations is
  higher than that measured by the WMAP satellite. 
If we were to repeat our simulations with the measured 
value of $f_{\rm b}$, we would expect all relations in
  Figure~\ref{fig:fgz1} to be shifted downwards, improving the
  agreement between our FO model and the observations.

However, we note that the observational mass estimates may be
  lower than the true mass, because they are derived from a $Y_{\rm
    X}-M_{500}$ relation that was calibrated using clusters with hydrostatic
  mass estimates. Simulations have shown that the assumption of
  hydrostatic equilibrium can bias such mass estimates low by $\sim
  10-20\%$ \citep{REM06,KDA07,NKV07,BHG08,PIV08,MRM09}, because of
  additional pressure support provided by subsonic bulk motions in the
  ICM and/or non-thermal components. This would imply a small
  systematic over-estimate of the gas fraction, so the
  observational data points in Figure~\ref{fig:fgz1} should be shifted
  downwards and to the right. Another potential source of systematic
  error is that the masses of high-redshift clusters in the sample of
  \citet{MJF08} were determined by assuming self-similar evolution of
  the $Y_{\rm X}-M_{500}$ relation.

It is also important to consider the effect of source
  selection on our results. Observational cluster selection is based
  on X-ray flux, so may be biased towards systems with higher baryon
  fractions, particularly at high redshift. It is not possible to
  quantify this effect precisely using the archival sample of
  \citet{MJF08} since their selection function is unknown. A simple
  way of estimating the impact of selection effects on our findings is
  to ask the question: given a typical flux-limited survey, which of
  our simulated clusters would actually be observed? For illustrative
  purposes, we choose a flux limit of $6.5\times
  10^{-14}$\,erg\,s$^{-1}$\,cm$^{-2}$, equal to that of the WARPS
  survey \citep{HPE08} from which many of the objects in the
  \citet{MJF08} sample are drawn. The effect of imposing this flux
  limit at $z=1$ is shown by the dashed line in Figure~\ref{fig:fgz1};
  only objects to the right of this line would be observed by a
  WARPS-like survey. Note that one of the clusters from \citet{MJF08} lies slightly to the left of this line; this is because it is at a lower redshift, $z\approx 0.8$. The important point to note is that, for both the
  PC and FO runs, there is only a narrow mass range where the bias is
  significant, with most clusters in the two samples remaining
  unaffected. Even if we took a much higher flux limit, it would
  require greatly increased scatter about the PC $f_{\rm gas}-M_{500}$
  relation for consistency with the observations, which is not
  intrinsic to the model.

\begin{figure}
\includegraphics[width=85mm]{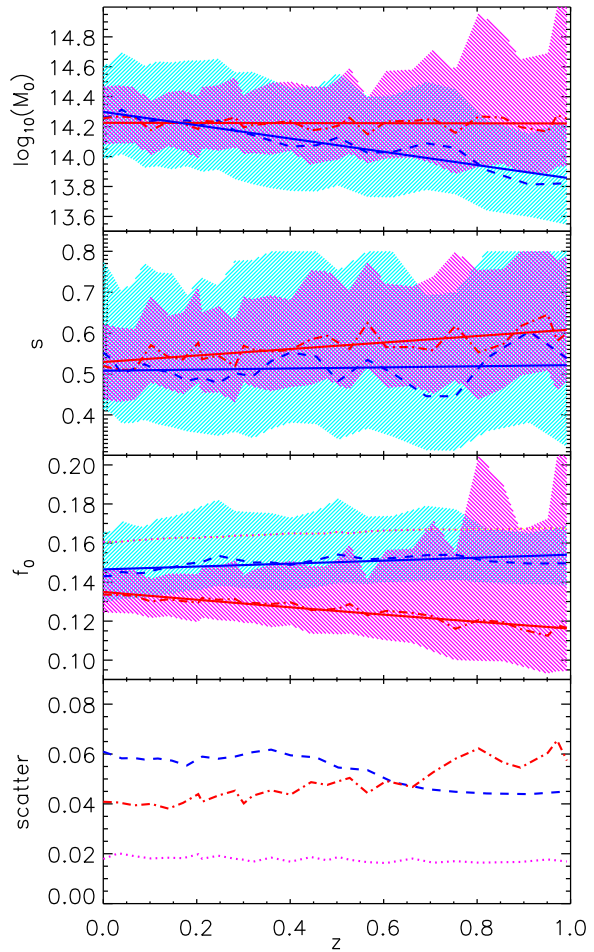}
\caption{Evolution of the fitting parameters of
  Equation~\ref{eq:fgmodel} for the gas fraction within $r_{500}$:
  \Nocool\ (dotted, magenta on yellow), \Precool\ (dash-dotted, red on
  magenta) and \Feedback\ (dashed, blue on cyan) lines.  The shaded
  region in each case shows the 1-sigma allowed parameter region.  The
  lowest panel shows the rms scatter in dex about the best-fit
  relation.  The solid lines in the upper three panels show the best fit
  linear relations to the redshift evolution of the parameters.}
\label{fig:fgfitevol}
\end{figure}

The best-fit parameters to the scaling relation of
Equation~\ref{eq:fgmodel} are shown as a function of redshift in
Figure~\ref{fig:fgfitevol}.  In the case of the \Nocool\ run, we fit only
the mean value of the gas fraction, $f_0$, which is well-determined.
For the other two runs, the shaded regions show the 1-sigma allowed
parameter range determined from monte-carlo markov chain fitting.  The
parameters show considerable scatter, but this scatter is highly
correlated.  So while it is formally possible for both the
\Precool\ and \Feedback\ clusters to have the same value of $f_0$ at
$z=1$, for example, the other parameters must adjust themselves so as
to maintain the difference in gas fraction seen in
Figure~\ref{fig:fgz1}.

The solid lines in the upper three panels of
Figure~\ref{fig:fgfitevol} show straight-line fits to the evolution of
each of the parameters with redshift -- note that we fit to
$\log_{10}f_0$ rather than $f_0$ as it is the former that appears in
Equation~\ref{eq:fgmodel}.  These fits are listed in
Table~\ref{tab:fgfitevol} and are used in the following section when
comparing our model predictions with observations.

\begin{table}
\begin{center}
  \caption{Best fit straight lines to the evolution with redshift of
    the gas fraction model parameters shown in
    Figure~\ref{fig:fgfitevol}.  These fits take the form
    $p=p_0+s_pz$, where $p$ is the parameter, $p_0$ its value at $z=0$,
    and $s_p$ the slope of the relation with redshift.}
\label{tab:fgfitevol}
\begin{tabular}{rrrr}
\hline
Parameter& Model& $p_0$& $s_p$\\
\hline
$m$& \Precool& 14.23& 0.00\\
   & \Feedback& 14.30& -0.44\\
$s$& \Precool& 0.53& 0.08\\
   & \Feedback& 0.51& 0.01\\
$\log_{10}{f_0}$& \Precool& -0.870& -0.065\\
               & \Feedback& -0.834& 0.022\\
\hline
\end{tabular}
\end{center}
\end{table}

\begin{figure}
\includegraphics[width=85mm]{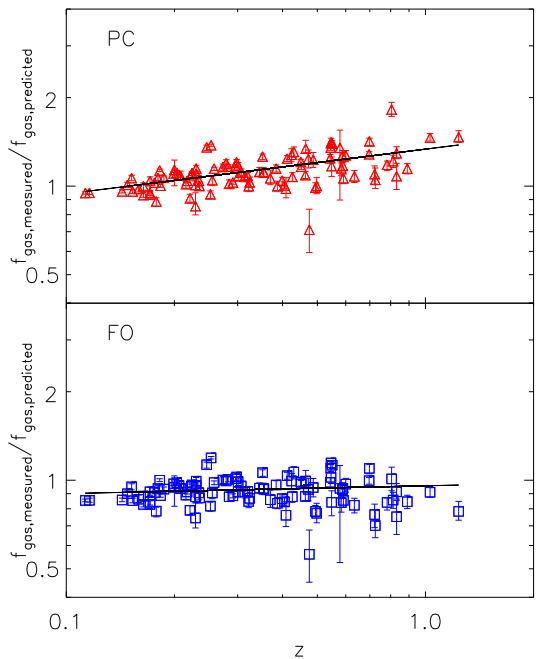}
\caption{Ratios of the observed hot gas fractions within $r_{500}$
  from \citet{MJF08} to our model predictions.  The
  solid lines show the best straight-line fits in $\log
  y$--$\log(1+z)$ space.}
\label{fig:fgevolobs}
\end{figure}

From the straight-line fits shown in Figure~\ref{fig:fgfitevol}, the mean gas
fraction within $r_{500}$ can be predicted for clusters of given mass
and redshift using Equation (\ref{eq:fgmodel}).  This prediction is compared 
to observed gas fractions from the sample of \citet{MJF08} in
Figure~\ref{fig:fgevolobs}. What is plotted here is the ratio of the observed
gas fraction to the predicted one, so that perfect agreement would
correspond to a value of unity, independent of redshift, but with some
scatter due to the cluster-to-cluster variation and measurement
error. The data are presented in this way since the gas fraction is
  a function of both cluster mass and redshift. The error
  bars are computed using the errors on the observed gas fractions
  (see above for details of how these were determined from the data of
  \citealt{MJF08}), accounting for the fact that the error on the
  total mass introduces an extra uncertainty when computing the
  theoretical prediction for the gas fraction.

It is immediately apparent that observations favour the
\Feedback\ prediction over the \Precool\ one, i.e.~limited evolution
in gas fraction since $z=1$.  This is shown in
Table~\ref{tab:fgevolobs} where we list the allowable parameter ranges
for straight line fits to the data both in linear and in $\log
y$--$\log(1+z)$ space, where $y$ is the ratio of observed to
predicted gas fraction within $r_{500}$.

\begin{table}
\begin{center}
  \caption{Best fit straight lines to the observed versus predicted
    hot gas ratios seen in Figure~\ref{fig:fgevolobs} in linear and in
  $\log y$--$\log(1+z)$ space.  The allowed 1-sigma
    parameter ranges are calculated assuming that the expected
    variance about the best fit is equal to the observed one.  The
    scatter is the root-mean square scatter about the best-fit line
    after allowing for the observational errors.  In the log-log
    plots, the scatter is expressed in dex.}
\label{tab:fgevolobs}
\begin{tabular}{rcccc}
\hline
Model& const& slope& scatter\\
\hline
 \Precool\ linear& $0.960\pm0.022$ & $0.436\pm0.064$ & 0.074\\
              log& $0.126\pm0.012$ & $0.152\pm0.020$ & 0.028\\
\Feedback\ linear& $0.923\pm0.019$ & $0.002\pm0.050$ & 0.071 \\ 
              log& $-0.019\pm0.012$ & $0.026\pm0.020$ & 0.034 \\
\hline
\end{tabular}
\end{center}
\end{table}

In making these fits, we treat the scatter about the mean relation as
an unknown, $\sigma_\mathrm{scatter}$, independent of mass and
redshift.  The data are not good enough for a more sophisticated model,
and that is likely, anyway, to make little difference to the fit.  Given
observational data, $y_i$, and errors, $\sigma_i$, we estimate the
scatter as
\begin{equation}
\sigma_\mathrm{scatter}^2={
{1\over N-2}\sum_i{(y_i-y_{i,\mathrm{fit}})^2\over\sigma_i^2}-1
\over {1\over N}\sum_i{1\over\sigma_i^2}},
\end{equation}
where $y_{i,\mathrm{fit}}$ are the best-fit values.  We iterate
to convergence in $\sigma_\mathrm{scatter}$, at each stage minimising
the chi-squared statistic
\begin{equation}
\chi^2=\sum_i{(y_i-y_{i,\mathrm{fit}})^2\over\sigma_i^2+\sigma_\mathrm{scatter}^2}.
\end{equation}

Note that the scatter about the best-fit line is, in each case, lower
than that seen in the simulations (as shown in the bottom panel of
Figure~\ref{fig:fgfitevol}).  Formally, therefore, neither of the
models is a good fit.  However, it seems unlikely that the true
scatter in $f_\mathrm{gas}$ will be below that seen in the
\Precool\ simulation. The uncertainty in the observed $f_\mathrm{gas}$
values is hard to determine, particularly at high-redshift, so it is
quite possible that the size of the error bars has been
over-estimated, leading to an under-estimate of the intrinsic scatter.

For the \Precool\ simulation, the slope of the observed to simulated
gas fraction ratio is incompatible with a horizontal line with high
significance. The difference between the best-fit values at
  $z=0.1$ and $z=1$ is about 6 times the scatter. Even if we were to
  account for observational bias in flux-limited samples towards
  clusters with higher baryon fractions, especially at high redshift,
  this is simply too large a difference to be explained by selection
  effects alone (recall our discussion of Figure~\ref{fig:fgz1}). We
conclude that the \Precool\ model can be ruled out as a viable cause
of entropy generation in the ICM.

The \Feedback\ simulations, on the other hand, are perfectly
consistent with a constant ratio of approximately unity.  The
  slightly lower mean hot gas fraction for the observations as
  compared to the simulations can be explained by the fact that the
  latter have a higher mean baryon fraction than the {\it WMAP}
  7-year value.5

We note that the analysis of \citet{EMT09} has many clusters
  in common with \citet{MJF08}, but lists total masses and gas
  fractions that are often in disagreement.  We are not certain why
  this is but note that there are differences in the analysis of the
  data, particularly in the cluster outskirts.  We have repeated the
  analysis described in this section with the data of \citet{EMT09}
  but the data are much less constraining, principally because they
  quote much larger error bars.  Nevertheless, one should bear in mind
  that systematic errors in the assumptions made in the data analysis
  could be degenerate with differences in the simulated ICM physics.

\begin{figure}
\includegraphics[width=85mm]{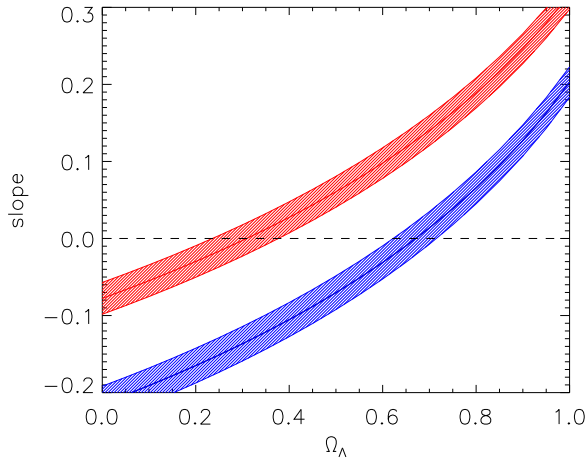}
\caption{The relationship between the slope in log-space of the
  observed/simulated gas fraction ratio as a function of redshift and
  the value of $\Omega_\Lambda$.  The upper, red curve is for the
  \Precool\ run and the lower, blue curve for the \Feedback\ run.  The
  shaded regions show the formal 1-sigma confidence regions.}
\label{fig:fgevolcosmo}
\end{figure}

Looking at the problem in reverse, one could ask what errors could be
introduced into the determination of cosmological parameters by the
choice of an incorrect physical model for the evolution of the ICM.
It is not possible to make a precise prediction for this using the
current simulations as we only have access to a single realisation
with a particular set of cosmological parameters.  Nevertheless,
fixing the simulated clusters to be the same,
Figure~\ref{fig:fgevolcosmo} shows the effect of changing the observed
cluster gas fractions in response to different values of
$\Omega_\Lambda$ (fixing $\Omega_\Lambda+\Omega_\mathrm{m}=1$).  From this, it can be seen that using an incorrect physical
model can have a dramatic effect, larger than that induced by changing
cosmological parameters within any reasonable range.

The analysis described in this section is necessarily very naive.  A
full treatment would require a detailed understanding of the selection
function of observed clusters, modelling of the scatter in the scaling
relations as a function of redshift, and of the mean relations as a
function of cosmological parameters.  Nevertheless, none of this is
likely to alter the basic conclusion that both observed clusters and
those derived from our \Feedback\ model show little evolution in hot
gas fractions within $r_{500}$ out to $z\approx 1$, whereas the
\Precool\ model predicts a strong decline.

\section{Conclusions}
\label{sec:conc}

In this paper, we investigate the baryon content of clusters of
galaxies in simulations using a variety of physical models for the
intracluster medium:
\begin{itemize}
\item GO -- gravitational heating only with no radiative cooling.  The
  purpose of this model is to test which aspects of the simulation
  evolve in a self-similar way and to provide a comparison for the
  other two runs.
\item PC - universal preheating to 200\,keV\,cm$^2$ at $z=4$, plus
  radiative cooling and star formation.  This represents widespread
  and early heating by objects that lie below the resolution limit of
  the simulation.
\item FO - feedback taken from a semi-analytic model, including
  heating from both supernovae and active galactic nuclei, but without
  radiative cooling.  The motivation for this model is to test heating
  from a realistic galaxy population that matches both the luminosity
  function and the black hole mass function of the current-day
  Universe.
\end{itemize}

The differential hot gas fraction profiles of clusters in the GO simulation are
approximately constant at radii greater than 0.2\,$r_{500}$, lying at
90 per cent of the cosmic mean.  In the other two simulations, the
profiles rise steeply from a low value in the cluster core before
bending over to an approximately constant value at large radii: for
the most massive clusters, $kT_\mathrm{sl}>5$\,keV, this occurs well
within $r_{500}$.

The cumulative hot gas fraction profiles of our clusters in both the PC and FO
runs lie well below those of the regular, cool-core (CC) clusters observed
by \citet{VKF06} and \citet{ARS08}.  However, they provide a fair
match to the non-cool-core (NCC) clusters found in the REXCESS
representative cluster survey \citep{PAP09}.

When we look at integrated gas fractions within fixed radii, the
agreement with observations is mixed.  The total gas fraction within
$r_{2500}$ shows a stronger variation with cluster mass in the
simulations than is seen in the observations.  On the other hand, the
agreement within $r_{500}$ is much better, at least on scales above
$5\times10^{14}\,h^{-1}\Msun$.  There is a small offset but that can
be explained by the fact that we adopt a mean baryon fraction in our
simulations that is higher than the current best-fit value from
WMAP, respectively 0.18 and 0.168.

A more slowly-varying function of mass is provided by the differential
gas fraction between $r_{2500}$ and $r_{500}$.  Unfortunately, the
scatter in the observational data is currently too large to allow any
meaningful comparison with the simulations.

In agreement with previous work, our simulated clusters show a much
smaller dependence of stellar fraction on mass than is seen in
observations.  Our stellar fractions within $r_{500}$ are about 0.013
for massive clusters, $M_{500}>10^{15}\,h^{-1}\Msun$, in the \Feedback\ run,
similar to observed values.  The \Precool\ run gives slightly higher
values, 0.02, whereas previous simulations can have stellar fractions
as high as 0.05 (i.e.\ as much as a third of all the baryons
within the cluster turned into stars).  On the other hand, for lower-mass
clusters, $M_{500}\sim5\times10^{13}\,h^{-1}\Msun$, our mean stellar fractions of
0.015 (\Feedback) and 0.03 (\Precool) are much lower than the observed
value of about 0.05.  We note that there is some theoretical
difficulty in understanding such a steep dependence of stellar
fraction on cluster mass, and that the observational determination of
this mass fraction is difficult, especially in low-mass systems.
While this should prove a fruitful line of investigation in the
future, it is probably too early to draw firm conclusions about the
validity of the models.

We have fitted the gas fractions as a function of mass to relations of
the form given by Equation~\ref{eq:fgmodel}, with the results shown in
Table~\ref{tab:fgfit}. The scatter about these mean relations is
lowest for the \Nocool\ run and significantly larger for the \Precool,
and especially the \Feedback\ runs.  Unfortunately, the observational
data are too poor to provide an accurate measure.

For the \Nocool\ run, we might expect that the scatter about the mean
gas fraction-mass relations has a physical origin in the formation
history of the clusters.  Indeed, there is a weak
correlation/anti-correlation of gas fraction with
concentration/formation time (most strongly with the expansion factor
at the time that the cluster had accumulated half its current mass).
The strongest correlation that might be used observationally to
correct for scatter in the gas fractions is that between deviation
from the mean $L_X$-$T_\mathrm{sl}$ relation and the excess gas fraction.

Although the gas fraction profiles are very similar for both the
\Precool\ and \Feedback\ runs at $z=0$, their evolution is very different.  In
the former gas is heated and expelled from the clusters at early
times, so that the gas is depleted at high redshift and gradually
falls back into the cluster over time.  By contrast, the continual
injection of energy in the \Feedback\ run leads to evolution that is
close to self-similar.

The evolution of halo gas fractions can therefore be used as a strong
discriminant between models.  We compare our simulated clusters with
the compilation of {\it Chandra} clusters from \citet{MJF08}.  The
observational data are fully-consistent with the
\Feedback\ predictions and disagree with the evolution seen in the
\Precool\ simulation with high significance.  We need to be a little
careful in interpreting these results as this is a highly biased
sample that may well contain a disproportionate number of luminous
clusters at high-redshift.  However, the scatter in gas fraction is
sufficiently small that, even if we only observed those clusters with
high baryon fractions at high redshift, the disagreement between the
observations and the \Precool\ prediction would still be significant.
We conclude that the observations
favour continual heating, as in our \Feedback\ model, over significant
preheating at high redshift.

A corollary of the strong dependence of gas fraction evolution on the
physics of entropy generation is that it becomes very difficult to use
the gas fraction as a probe of cosmology.  The differences caused by
uncertain gas physics currently swamp those caused by
reasonable changes in cosmological parameters.  In the
  future, as both observational data for high-redshift clusters and
  models of the ICM improve, a joint analysis
  should be undertaken that considers variations in both cosmological
  parameters and cluster physics.

The main limitation to our present study is that the absence of
cooling in our \Feedback\ simulation leaves us unable to model CC
clusters.  Whilst that does not significantly affect the gas fractions
when integrated out to $r_{500}$, it would be clearly desirable to also
reproduce the full range of gas fraction profiles at smaller radii.
We are working on ways to introduce cooling into the \Feedback\ scheme
without leading to excess production of cooled gas in CC clusters.

\section*{Acknowledgements}
This work was supported by the STFC [grant number ST/F002858/1, PhD studentship]\\
 We would like to thank Adrian Jenkins for supplying the
code to generate the initial conditions.  The simulations were carried
out using the HPC facility at the University of Nottingham and the
Virgo Consortium facilities at Durham.  Analysis was undertaken on the
Archimedes cluster at Sussex.  We are grateful to Steve Allen, Stefano
Andreon, Gabriel Pratt, Alistair Sanderson, Alexey Vikhlinin and
especially Ben Maughan for supplying us with observational data and
helping us with its interpretation.
We would like to thank the referee for their constructive 
comments which have strengthened this paper.

\bibliographystyle{mn2e}
\bibliography{bibliography}

\begin{thebibliography}{75}
\expandafter\ifx\csname natexlab\endcsname\relax\def\natexlab#1{#1}\fi

\bibitem[{{Allen} {et~al.}(2008){Allen}, {Rapetti}, {Schmidt}, {Ebeling},
  {Morris}, \& {Fabian}}]{ARS08}
{Allen} S.~W., {Rapetti} D.~A., {Schmidt} R.~W., {Ebeling} H., {Morris} R.~G.,
  {Fabian} A.~C., 2008, MNRAS, 383, 879

\bibitem[{{Allen} {et~al.}(2004){Allen}, {Schmidt}, {Ebeling}, {Fabian}, \&
  {van Speybroeck}}]{ASE04}
{Allen} S.~W., {Schmidt} R.~W., {Ebeling} H., {Fabian} A.~C., {van Speybroeck}
  L., 2004, MNRAS, 353, 457

\bibitem[{{Allen} {et~al.}(2002){Allen}, {Schmidt}, \& {Fabian}}]{ASF02}
{Allen} S.~W., {Schmidt} R.~W., {Fabian} A.~C., 2002, MNRAS, 334, L11

\bibitem[{{Andreon}(2010)}]{And10}
{Andreon} S., 2010, ArXiv e-prints: 1004.2785

\bibitem[{{Arnaud} {et~al.}(2007){Arnaud}, {Pointecouteau}, \& {Pratt}}]{APP07}
{Arnaud} M., {Pointecouteau} E., {Pratt} G.~W., 2007, A\&A, 474, L37

\bibitem[{{Ascasibar} {et~al.}(2006){Ascasibar}, {Sevilla}, {Yepes},
  {M{\"u}ller}, \& {Gottl{\"o}ber}}]{ASY06}
{Ascasibar} Y., {Sevilla} R., {Yepes} G., {M{\"u}ller} V., {Gottl{\"o}ber} S.,
  2006, MNRAS, 371, 193

\bibitem[{{Balogh} {et~al.}(2008){Balogh}, {McCarthy}, {Bower}, \&
  {Eke}}]{BMB08a}
{Balogh} M.~L., {McCarthy} I.~G., {Bower} R.~G., {Eke} V.~R., 2008, MNRAS, 385,
  1003

\bibitem[{{Bernstein} {et~al.}(1995){Bernstein}, {Nichol}, {Tyson}, {Ulmer}, \&
  {Wittman}}]{BNT95}
{Bernstein} G.~M., {Nichol} R.~C., {Tyson} J.~A., {Ulmer} M.~P., {Wittman} D.,
  1995, AJ, 110, 1507

\bibitem[{{Bialek} {et~al.}(2001){Bialek}, {Evrard}, \& {Mohr}}]{BEM01}
{Bialek} J.~J., {Evrard} A.~E., {Mohr} J.~J., 2001, ApJ, 555, 597

\bibitem[{{B{\"o}hringer} {et~al.}(2007){B{\"o}hringer}, {Schuecker}, {Pratt},
  {Arnaud}, {Ponman}, {Croston}, {Borgani}, {Bower}, {Briel}, {Collins},
  {Donahue}, {Forman}, {Finoguenov}, {Geller}, {Guzzo}, {Henry}, {Kneissl},
  {Mohr}, {Matsushita}, {Mullis}, {Ohashi}, {Pedersen}, {Pierini}, {Quintana},
  {Raychaudhury}, {Reiprich}, {Romer}, {Rosati}, {Sabirli}, {Temple}, {Viana},
  {Vikhlinin}, {Voit}, \& {Zhang}}]{BSP07}
{B{\"o}hringer} H., {Schuecker} P., {Pratt} G.~W., {Arnaud} M., {Ponman} T.~J.,
  {Croston} J.~H., {Borgani} S., {Bower} R.~G., {Briel} U.~G., {Collins} C.~A.,
  {Donahue} M., {Forman} W.~R., {Finoguenov} A., {Geller} M.~J., {Guzzo} L.,
  {Henry} J.~P., {Kneissl} R., {Mohr} J.~J., {Matsushita} K., {Mullis} C.~R.,
  {Ohashi} T., {Pedersen} K., {Pierini} D., {Quintana} H., {Raychaudhury} S.,
  {Reiprich} T.~H., {Romer} A.~K., {Rosati} P., {Sabirli} K., {Temple} R.~F.,
  {Viana} P.~T.~P., {Vikhlinin} A., {Voit} G.~M., {Zhang} Y., 2007, A\&A, 469,
  363

\bibitem[{{Borgani} {et~al.}(2005){Borgani}, {Finoguenov}, {Kay}, {Ponman},
  {Springel}, {Tozzi}, \& {Voit}}]{BFK05}
{Borgani} S., {Finoguenov} A., {Kay} S.~T., {Ponman} T.~J., {Springel} V.,
  {Tozzi} P., {Voit} G.~M., 2005, MNRAS, 361, 233

\bibitem[{{Borgani} {et~al.}(2002){Borgani}, {Governato}, {Wadsley}, {Menci},
  {Tozzi}, {Quinn}, {Stadel}, \& {Lake}}]{BGW02}
{Borgani} S., {Governato} F., {Wadsley} J., {Menci} N., {Tozzi} P., {Quinn} T.,
  {Stadel} J., {Lake} G., 2002, MNRAS, 336, 409

\bibitem[{{Borgani} {et~al.}(2004){Borgani}, {Murante}, {Springel}, {Diaferio},
  {Dolag}, {Moscardini}, {Tormen}, {Tornatore}, \& {Tozzi}}]{BMS04}
{Borgani} S., {Murante} G., {Springel} V., {Diaferio} A., {Dolag} K.,
  {Moscardini} L., {Tormen} G., {Tornatore} L., {Tozzi} P., 2004, MNRAS, 348,
  1078

\bibitem[{{Brighenti} \& {Mathews}(2001)}]{BRM01}
{Brighenti} F., {Mathews} W.~G., 2001, ApJ, 553, 103

\bibitem[{{Burns} {et~al.}(2008){Burns}, {Hallman}, {Gantner}, {Motl}, \&
  {Norman}}]{BHG08}
{Burns} J.~O., {Hallman} E.~J., {Gantner} B., {Motl} P.~M., {Norman} M.~L.,
  2008, ApJ, 675, 1125

\bibitem[{{Colless} {et~al.}(2001){Colless}, {Dalton}, {Maddox}, {Sutherland},
  {Norberg}, {Cole}, {Bland-Hawthorn}, {Bridges}, {Cannon}, {Collins}, {Couch},
  {Cross}, {Deeley}, {De Propris}, {Driver}, {Efstathiou}, {Ellis}, {Frenk},
  {Glazebrook}, {Jackson}, {Lahav}, {Lewis}, {Lumsden}, {Madgwick}, {Peacock},
  {Peterson}, {Price}, {Seaborne}, \& {Taylor}}]{CDM01}
{Colless} M., {Dalton} G., {Maddox} S., {Sutherland} W., {Norberg} P., {Cole}
  S., {Bland-Hawthorn} J., {Bridges} T., {Cannon} R., {Collins} C., {Couch} W.,
  {Cross} N., {Deeley} K., {De Propris} R., {Driver} S.~P., {Efstathiou} G.,
  {Ellis} R.~S., {Frenk} C.~S., {Glazebrook} K., {Jackson} C., {Lahav} O.,
  {Lewis} I., {Lumsden} S., {Madgwick} D., {Peacock} J.~A., {Peterson} B.~A.,
  {Price} I., {Seaborne} M., {Taylor} K., 2001, MNRAS, 328, 1039

\bibitem[{{Crain} {et~al.}(2007){Crain}, {Eke}, {Frenk}, {Jenkins}, {McCarthy},
  {Navarro}, \& {Pearce}}]{CEF07}
{Crain} R.~A., {Eke} V.~R., {Frenk} C.~S., {Jenkins} A., {McCarthy} I.~G.,
  {Navarro} J.~F., {Pearce} F.~R., 2007, MNRAS, 377, 41

\bibitem[{{Dav{\'e}} {et~al.}(2008){Dav{\'e}}, {Oppenheimer}, \&
  {Sivanandam}}]{DOS08}
{Dav{\'e}} R., {Oppenheimer} B.~D., {Sivanandam} S., 2008, MNRAS, 391, 110

\bibitem[{{De Lucia} \& {Blaizot}(2007)}]{LuB07}
{De Lucia} G., {Blaizot} J., 2007, MNRAS, 375, 2

\bibitem[{{Eisenhardt} {et~al.}(2008){Eisenhardt}, {Brodwin}, {Gonzalez},
  {Stanford}, {Stern}, {Barmby}, {Brown}, {Dawson}, {Dey}, {Doi}, {Galametz},
  {Jannuzi}, {Kochanek}, {Meyers}, {Morokuma}, \& {Moustakas}}]{EBG08}
{Eisenhardt} P.~R.~M., {Brodwin} M., {Gonzalez} A.~H., {Stanford} S.~A.,
  {Stern} D., {Barmby} P., {Brown} M.~J.~I., {Dawson} K., {Dey} A., {Doi} M.,
  {Galametz} A., {Jannuzi} B.~T., {Kochanek} C.~S., {Meyers} J., {Morokuma} T.,
  {Moustakas} L.~A., 2008, ApJ, 684, 905

\bibitem[{{Eke} {et~al.}(1998){Eke}, {Navarro}, \& {Frenk}}]{ENF98}
{Eke} V.~R., {Navarro} J.~F., {Frenk} C.~S., 1998, ApJ, 503, 569

\bibitem[{{Ettori} {et~al.}(2006){Ettori}, {Dolag}, {Borgani}, \&
  {Murante}}]{EDB06}
{Ettori} S., {Dolag} K., {Borgani} S., {Murante} G., 2006, MNRAS, 365, 1021

\bibitem[{{Ettori} {et~al.}(2009){Ettori}, {Morandi}, {Tozzi}, {Balestra},
  {Borgani}, {Rosati}, {Lovisari}, \& {Terenziani}}]{EMT09}
{Ettori} S., {Morandi} A., {Tozzi} P., {Balestra} I., {Borgani} S., {Rosati}
  P., {Lovisari} L., {Terenziani} F., 2009, A\&A, 501, 61

\bibitem[{{Evrard} \& {Henry}(1991)}]{EVH91}
{Evrard} A.~E., {Henry} J.~P., 1991, ApJ, 383, 95

\bibitem[{{Fabjan} {et~al.}(2010){Fabjan}, {Borgani}, {Tornatore}, {Saro},
  {Murante}, \& {Dolag}}]{FBT10}
{Fabjan} D., {Borgani} S., {Tornatore} L., {Saro} A., {Murante} G., {Dolag} K.,
  2010, MNRAS, 401

\bibitem[{{Feldmeier} {et~al.}(2004){Feldmeier}, {Mihos}, {Morrison},
  {Harding}, {Kaib}, \& {Dubinski}}]{FMM04}
{Feldmeier} J.~J., {Mihos} J.~C., {Morrison} H.~L., {Harding} P., {Kaib} N.,
  {Dubinski} J., 2004, ApJ, 609, 617

\bibitem[{{Feldmeier} {et~al.}(2002){Feldmeier}, {Mihos}, {Morrison}, {Rodney},
  \& {Harding}}]{FMM02}
{Feldmeier} J.~J., {Mihos} J.~C., {Morrison} H.~L., {Rodney} S.~A., {Harding}
  P., 2002, ApJ, 575, 779

\bibitem[{{Giodini} {et~al.}(2009){Giodini}, {Pierini}, {Finoguenov}, {Pratt},
  {Boehringer}, {Leauthaud}, {Guzzo}, {Aussel}, \& {the COSMOS
  collaboration}}]{GPF09}
{Giodini} S., {Pierini} D., {Finoguenov} A., {Pratt} G.~W., {Boehringer} H.,
  {Leauthaud} A., {Guzzo} L., {Aussel} H., {the COSMOS collaboration}, 2009,
  ApJ, 703, 982

\bibitem[{{Gonzalez} {et~al.}(2005){Gonzalez}, {Zabludoff}, \&
  {Zaritsky}}]{GZZ05}
{Gonzalez} A.~H., {Zabludoff} A.~I., {Zaritsky} D., 2005, ApJ, 618, 195

\bibitem[{{Gonzalez} {et~al.}(2000){Gonzalez}, {Zabludoff}, {Zaritsky}, \&
  {Dalcanton}}]{GZZ00}
{Gonzalez} A.~H., {Zabludoff} A.~I., {Zaritsky} D., {Dalcanton} J.~J., 2000,
  ApJ, 536, 561

\bibitem[{{Gonzalez} {et~al.}(2007){Gonzalez}, {Zaritsky}, \&
  {Zabludoff}}]{GZZ07}
{Gonzalez} A.~H., {Zaritsky} D., {Zabludoff} A.~I., 2007, ApJ, 666, 147

\bibitem[{{Horner} {et~al.}(2008){Horner}, {Perlman}, {Ebeling}, {Jones},
  {Scharf}, {Wegner}, {Malkan}, \& {Maughan}}]{HPE08}
{Horner} D.~J., {Perlman} E.~S., {Ebeling} H., {Jones} L.~R., {Scharf} C.~A.,
  {Wegner} G., {Malkan} M., {Maughan} B., 2008, ApJ Supp., 176, 374

\bibitem[{{Kaiser}(1991)}]{KAI91}
{Kaiser} N., 1991, ApJ, 383, 104

\bibitem[{{Kay} {et~al.}(2007{\natexlab{a}}){Kay}, {da Silva}, {Aghanim},
  {Blanchard}, {Liddle}, {Puget}, {Sadat}, \& {Thomas}}]{KSA07}
{Kay} S.~T., {da Silva} A.~C., {Aghanim} N., {Blanchard} A., {Liddle} A.~R.,
  {Puget} J.-L., {Sadat} R., {Thomas} P.~A., 2007{\natexlab{a}}, MNRAS, 377,
  317

\bibitem[{{Kay} {et~al.}(2007{\natexlab{b}}){Kay}, {da Silva}, {Aghanim},
  {Blanchard}, {Liddle}, {Puget}, {Sadat}, \& {Thomas}}]{KDA07}
---, 2007{\natexlab{b}}, MNRAS, 377, 317

\bibitem[{{Kitzbichler} \& {White}(2008)}]{KiW08}
{Kitzbichler} M.~G., {White} S.~D.~M., 2008, MNRAS, 391, 1489

\bibitem[{{Komatsu} {et~al.}(2010){Komatsu}, {Smith}, {Dunkley}, {Bennett},
  {Gold}, {Hinshaw}, {Jarosik}, {Larson}, {Nolta}, {Page}, {Spergel},
  {Halpern}, {Hill}, {Kogut}, {Limon}, {Meyer}, {Odegard}, {Tucker}, {Weiland},
  {Wollack}, \& {Wright}}]{KSD10}
{Komatsu} E., {Smith} K.~M., {Dunkley} J., {Bennett} C.~L., {Gold} B.,
  {Hinshaw} G., {Jarosik} N., {Larson} D., {Nolta} M.~R., {Page} L., {Spergel}
  D.~N., {Halpern} M., {Hill} R.~S., {Kogut} A., {Limon} M., {Meyer} S.~S.,
  {Odegard} N., {Tucker} G.~S., {Weiland} J.~L., {Wollack} E., {Wright} E.~L.,
  2010, ArXiv e-prints: 1001.4538

\bibitem[{{Kravtsov} {et~al.}(2005){Kravtsov}, {Nagai}, \& {Vikhlinin}}]{KNV05}
{Kravtsov} A.~V., {Nagai} D., {Vikhlinin} A.~A., 2005, ApJ, 625, 588

\bibitem[{{Krick} {et~al.}(2006){Krick}, {Bernstein}, \& {Pimbblet}}]{KBP06}
{Krick} J.~E., {Bernstein} R.~A., {Pimbblet} K.~A., 2006, ApJ, 131, 168

\bibitem[{{Lagan{\'a}} {et~al.}(2008){Lagan{\'a}}, {Lima Neto},
  {Andrade-Santos}, \& {Cypriano}}]{LLA08}
{Lagan{\'a}} T.~F., {Lima Neto} G.~B., {Andrade-Santos} F., {Cypriano} E.~S.,
  2008, A\&A, 485, 633

\bibitem[{{LaRoque} {et~al.}(2006){LaRoque}, {Bonamente}, {Carlstrom}, {Joy},
  {Nagai}, {Reese}, \& {Dawson}}]{LBC06}
{LaRoque} S.~J., {Bonamente} M., {Carlstrom} J.~E., {Joy} M.~K., {Nagai} D.,
  {Reese} E.~D., {Dawson} K.~S., 2006, ApJ, 652, 917

\bibitem[{{Mantz} {et~al.}(2009{\natexlab{a}}){Mantz}, {Allen}, {Ebeling},
  {Rapetti}, \& {Drlica-Wagner}}]{MAE09}
{Mantz} A., {Allen} S.~W., {Ebeling} H., {Rapetti} D., {Drlica-Wagner} A.,
  2009{\natexlab{a}}, ArXiv e-prints: 0909.3099

\bibitem[{{Mantz} {et~al.}(2009{\natexlab{b}}){Mantz}, {Allen}, {Rapetti}, \&
  {Ebeling}}]{MAR09}
{Mantz} A., {Allen} S.~W., {Rapetti} D., {Ebeling} H., 2009{\natexlab{b}},
  ArXiv e-prints: 0909.3098

\bibitem[{{Maughan} {et~al.}(2008){Maughan}, {Jones}, {Forman}, \& {Van
  Speybroeck}}]{MJF08}
{Maughan} B.~J., {Jones} C., {Forman} W., {Van Speybroeck} L., 2008, ApJ Supp.,
  174, 117

\bibitem[{{Mazzotta} {et~al.}(2004){Mazzotta}, {Rasia}, {Moscardini}, \&
  {Tormen}}]{MRM04}
{Mazzotta} P., {Rasia} E., {Moscardini} L., {Tormen} G., 2004, MNRAS, 354, 10

\bibitem[{{McCarthy} {et~al.}(2009){McCarthy}, {Schaye}, {Ponman}, {Bower},
  {Booth}, {Dalla Vecchia}, {Crain}, {Springel}, {Theuns}, \&
  {Wiersma}}]{CSP09}
{McCarthy} I.~G., {Schaye} J., {Ponman} T.~J., {Bower} R.~G., {Booth} C.~M.,
  {Dalla Vecchia} C., {Crain} R.~A., {Springel} V., {Theuns} T., {Wiersma}
  R.~P.~C., 2009, ArXiv e-prints: 0911.2641

\bibitem[{{Meneghetti} {et~al.}(2009){Meneghetti}, {Rasia}, {Merten},
  {Bellagamba}, {Ettori}, {Mazzotta}, \& {Dolag}}]{MRM09}
{Meneghetti} M., {Rasia} E., {Merten} J., {Bellagamba} F., {Ettori} S.,
  {Mazzotta} P., {Dolag} K., 2009, arXiv: astro-ph/0912.1343

\bibitem[{{Mo} {et~al.}(2005){Mo}, {Yang}, {van den Bosch}, \& {Katz}}]{MYB05}
{Mo} H.~J., {Yang} X., {van den Bosch} F.~C., {Katz} N., 2005, MNRAS, 363, 1155

\bibitem[{{Muanwong} {et~al.}(2006){Muanwong}, {Kay}, \& {Thomas}}]{MKT06}
{Muanwong} O., {Kay} S.~T., {Thomas} P.~A., 2006, ApJ, 649, 640

\bibitem[{{Muanwong} {et~al.}(2002){Muanwong}, {Thomas}, {Kay}, \&
  {Pearce}}]{MTK02}
{Muanwong} O., {Thomas} P.~A., {Kay} S.~T., {Pearce} F.~R., 2002, MNRAS, 336,
  527

\bibitem[{{Nagai} {et~al.}(2007){Nagai}, {Kravtsov}, \& {Vikhlinin}}]{NKV07}
{Nagai} D., {Kravtsov} A.~V., {Vikhlinin} A., 2007, ApJ, 668, 1

\bibitem[{{Navarro} {et~al.}(1995){Navarro}, {Frenk}, \& {White}}]{NFW95}
{Navarro} J.~F., {Frenk} C.~S., {White} S.~D.~M., 1995, MNRAS, 275, 720

\bibitem[{{Navarro} {et~al.}(1997){Navarro}, {Frenk}, \& {White}}]{NFW97}
---, 1997, ApJ, 490, 493

\bibitem[{{Pearce} {et~al.}(1994){Pearce}, {Thomas}, \& {Couchman}}]{PTC94}
{Pearce} F.~R., {Thomas} P.~A., {Couchman} H.~M.~P., 1994, MNRAS, 268, 953

\bibitem[{{Piffaretti} \& {Valdarnini}(2008)}]{PIV08}
{Piffaretti} R., {Valdarnini} R., 2008, A\&A, 491, 71

\bibitem[{{Ponman} {et~al.}(2003){Ponman}, {Sanderson}, \&
  {Finoguenov}}]{PSF03}
{Ponman} T.~J., {Sanderson} A.~J.~R., {Finoguenov} A., 2003, MNRAS, 343, 331

\bibitem[{{Pratt} {et~al.}(2010){Pratt}, {Arnaud}, {Piffaretti},
  {B{\"o}hringer}, {Ponman}, {Croston}, {Voit}, {Borgani}, \& {Bower}}]{PAP09}
{Pratt} G.~W., {Arnaud} M., {Piffaretti} R., {B{\"o}hringer} H., {Ponman}
  T.~J., {Croston} J.~H., {Voit} G.~M., {Borgani} S., {Bower} R.~G., 2010,
  A\&A, 511, A85+

\bibitem[{{Pratt} {et~al.}(2006){Pratt}, {Arnaud}, \& {Pointecouteau}}]{PAP06}
{Pratt} G.~W., {Arnaud} M., {Pointecouteau} E., 2006, A\&A, 446, 429

\bibitem[{{Puchwein} {et~al.}(2010){Puchwein}, {Springel}, {Sijacki}, \&
  {Dolag}}]{PSS10}
{Puchwein} E., {Springel} V., {Sijacki} D., {Dolag} K., 2010, ArXiv e-prints:
  1001.3018

\bibitem[{{Rasia} {et~al.}(2006){Rasia}, {Ettori}, {Moscardini}, {Mazzotta},
  {Borgani}, {Dolag}, {Tormen}, {Cheng}, \& {Diaferio}}]{REM06}
{Rasia} E., {Ettori} S., {Moscardini} L., {Mazzotta} P., {Borgani} S., {Dolag}
  K., {Tormen} G., {Cheng} L.~M., {Diaferio} A., 2006, MNRAS, 369, 2013

\bibitem[{{Sadat} {et~al.}(2005){Sadat}, {Blanchard}, {Vauclair}, {Lumb},
  {Bartlett}, {Romer}, {Bernard}, {Boer}, {Marty}, {Nevalainen}, {Burke},
  {Collins}, \& {Nichol}}]{SBV05}
{Sadat} R., {Blanchard} A., {Vauclair} S.~C., {Lumb} D.~H., {Bartlett} J.,
  {Romer} A.~K., {Bernard} J., {Boer} M., {Marty} P., {Nevalainen} J., {Burke}
  D.~J., {Collins} C.~A., {Nichol} R.~C., 2005, A\&A, 437, 31

\bibitem[{{Sanderson} {et~al.}(2003){Sanderson}, {Ponman}, {Finoguenov},
  {Lloyd-Davies}, \& {Markevitch}}]{SPF03}
{Sanderson} A.~J.~R., {Ponman} T.~J., {Finoguenov} A., {Lloyd-Davies} E.~J.,
  {Markevitch} M., 2003, MNRAS, 340, 989

\bibitem[{{Short} \& {Thomas}(2009)}]{SHT09}
{Short} C.~J., {Thomas} P.~A., 2009, ApJ, 704, 915

\bibitem[{Short {et~al.}(2010)Short, Thomas, Young, Pearce, Jenkins, \&
  Muanwong}]{STY10}
Short C.~J., Thomas P.~A., Young O.~E., Pearce F.~R., Jenkins A., Muanwong O.,
  2010, ArXiv e-prints: 1002.4539

\bibitem[{{Spergel} {et~al.}(2003){Spergel}, {Verde}, {Peiris}, {Komatsu},
  {Nolta}, {Bennett}, {Halpern}, {Hinshaw}, {Jarosik}, {Kogut}, {Limon},
  {Meyer}, {Page}, {Tucker}, {Weiland}, {Wollack}, \& {Wright}}]{SVP03}
{Spergel} D.~N., {Verde} L., {Peiris} H.~V., {Komatsu} E., {Nolta} M.~R.,
  {Bennett} C.~L., {Halpern} M., {Hinshaw} G., {Jarosik} N., {Kogut} A.,
  {Limon} M., {Meyer} S.~S., {Page} L., {Tucker} G.~S., {Weiland} J.~L.,
  {Wollack} E., {Wright} E.~L., 2003, ApJ Supp., 148, 175

\bibitem[{{Springel} {et~al.}(2005){Springel}, {White}, {Jenkins}, {Frenk},
  {Yoshida}, {Gao}, {Navarro}, {Thacker}, {Croton}, {Helly}, {Peacock}, {Cole},
  {Thomas}, {Couchman}, {Evrard}, {Colberg}, \& {Pearce}}]{SWJ05}
{Springel} V., {White} S.~D.~M., {Jenkins} A., {Frenk} C.~S., {Yoshida} N.,
  {Gao} L., {Navarro} J., {Thacker} R., {Croton} D., {Helly} J., {Peacock}
  J.~A., {Cole} S., {Thomas} P., {Couchman} H., {Evrard} A., {Colberg} J.,
  {Pearce} F., 2005, Nat., 435, 629

\bibitem[{{Stanek} {et~al.}(2010){Stanek}, {Rasia}, {Evrard}, {Pearce}, \&
  {Gazzola}}]{SRE09}
{Stanek} R., {Rasia} E., {Evrard} A.~E., {Pearce} F., {Gazzola} L., 2010, ApJ,
  715, 1508

\bibitem[{{Sun} {et~al.}(2009){Sun}, {Voit}, {Donahue}, {Jones}, {Forman}, \&
  {Vikhlinin}}]{SVD09}
{Sun} M., {Voit} G.~M., {Donahue} M., {Jones} C., {Forman} W., {Vikhlinin} A.,
  2009, ApJ, 693, 1142

\bibitem[{{Sutherland} \& {Dopita}(1993)}]{SUD93}
{Sutherland} R.~S., {Dopita} M.~A., 1993, ApJ Supp., 88, 253

\bibitem[{{Tornatore} {et~al.}(2003){Tornatore}, {Borgani}, {Springel},
  {Matteucci}, {Menci}, \& {Murante}}]{TBS03}
{Tornatore} L., {Borgani} S., {Springel} V., {Matteucci} F., {Menci} N.,
  {Murante} G., 2003, MNRAS, 342, 1025

\bibitem[{{Tozzi} {et~al.}(2003){Tozzi}, {Rosati}, {Ettori}, {Borgani},
  {Mainieri}, \& {Norman}}]{TRE03}
{Tozzi} P., {Rosati} P., {Ettori} S., {Borgani} S., {Mainieri} V., {Norman} C.,
  2003, ApJ, 593, 705

\bibitem[{{Vikhlinin} {et~al.}(2006){Vikhlinin}, {Kravtsov}, {Forman}, {Jones},
  {Markevitch}, {Murray}, \& {Van Speybroeck}}]{VKF06}
{Vikhlinin} A., {Kravtsov} A., {Forman} W., {Jones} C., {Markevitch} M.,
  {Murray} S.~S., {Van Speybroeck} L., 2006, ApJ, 640, 691

\bibitem[{{Voit} {et~al.}(2005){Voit}, {Kay}, \& {Bryan}}]{VKB05}
{Voit} G.~M., {Kay} S.~T., {Bryan} G.~L., 2005, MNRAS, 364, 909

\bibitem[{{Weiner} {et~al.}(2009){Weiner}, {Coil}, {Prochaska}, {Newman},
  {Cooper}, {Bundy}, {Conselice}, {Dutton}, {Faber}, {Koo}, {Lotz}, {Rieke}, \&
  {Rubin}}]{WCP09}
{Weiner} B.~J., {Coil} A.~L., {Prochaska} J.~X., {Newman} J.~A., {Cooper}
  M.~C., {Bundy} K., {Conselice} C.~J., {Dutton} A.~A., {Faber} S.~M., {Koo}
  D.~C., {Lotz} J.~M., {Rieke} G.~H., {Rubin} K.~H.~R., 2009, ApJ, 692, 187

\bibitem[{{Zibetti} {et~al.}(2005){Zibetti}, {White}, {Schneider}, \&
  {Brinkmann}}]{ZWS05}
{Zibetti} S., {White} S.~D.~M., {Schneider} D.~P., {Brinkmann} J., 2005, MNRAS,
  358, 949

\end{thebibliography}

\section*{Appendix}

\begin{figure}
\includegraphics[width=85mm]{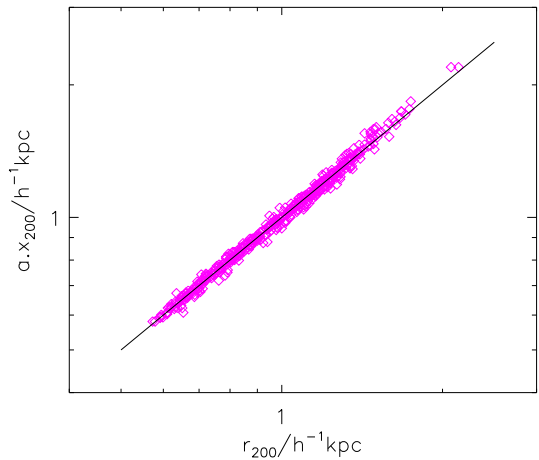}
\caption{The abscissa shows the actual value of $r_{200}$ for each of
  the clusters, whilst the ordinate shows the value predicted from
  $M_{2500}$ and $M_{500}$ assuming an NFW profile.}
\label{fig:nfwtest}
\end{figure}

The central concentration of clusters can be determined by measuring
the mass at two different over-densities, for example $M_{2500}$ and
$M_{500}$.  Together, these uniquely determine the parameters of the
NFW profile \citep{NFW97} without any need to fit the profile as a
function of radius.\footnote{We provide IDL routines to do this at
  {\tt http://astronomy.susx.ac.uk/$\sim$petert/nfw.pro.}}  To test the
effectiveness of this procedure, we show in Figure~\ref{fig:nfwtest}
the value of $r_{200}$ predicted by the NFW fit (i.e.~the NFW scale radius,
$a$, times the concentration, $x_{200}$) versus the actual 
value.  As can be seen, the two agree very well, confirming that the
clusters are well-fit by the NFW profile out to this radius.

\begin{figure}
\includegraphics[width=85mm]{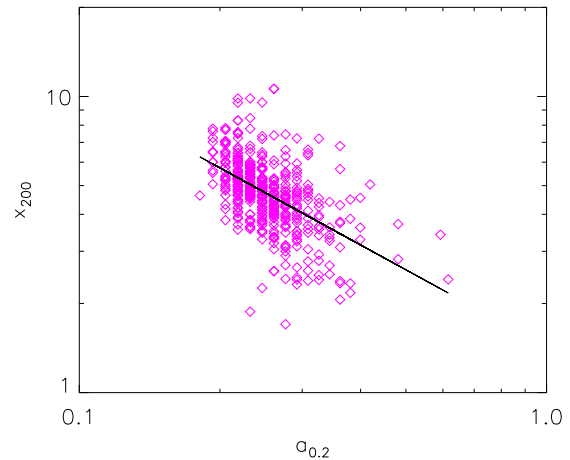}
\caption{Cluster concentration versus the expansion factor at the time
  that the clusters have accumulated one fifth of their final mass.
  The solid line shows the best-fitting power-law correlation.}
\label{fig:conc}
\end{figure}

Cluster concentrations are often thought to have a dependence upon
cluster mass, with more massive clusters having lower concentrations.
That is indeed the case, but we find a much stronger correlation with
cluster formation time, as illustrated in Figure~\ref{fig:conc}.  Here
the formation time is taken to be the value of the expansion factor
when the total mass of all the subhalos that will go on to make up the
cluster equal one fifth of the final cluster mass, but other
definitions give similar correlations.  We plot expansion factor
rather than age as this gives a more linear correlation.  The results
are shown here for the \Nocool\ simulation; those for the
\Precool\ and \Feedback\ runs are very slightly different because of
the contribution of the baryons to the total mass.

On removing the best-fitting correlation (shown as a solid line in the
figure), the residual concentration shows no dependence on mass.
The correlation with mass is thus a secondary one that follows because
low mass clusters tend to form at lower expansion factors than more
massive ones.

\label{lastpage}

\end{document}